\documentclass[preprint2]{aastex6}
\pdfoutput=1

\usepackage{float}
\usepackage{hyperref}
\usepackage{amsmath}
\usepackage{booktabs}
\usepackage{longtable}
\usepackage{xspace}
\usepackage[T1]{fontenc}

\newcommand{\Kepler}{\textit{Kepler}\xspace}

\newcommand{\isochrones}{\texttt{isochrones}\xspace}
\newcommand{\SpecMatch}{\texttt{SpecMatch-Emp}\xspace}
\newcommand{\Gaia}{\textit{Gaia}\xspace}
\newcommand{\Hipparcos}{\textit{Hipparcos}\xspace}
\newcommand{\chisq}{\ensuremath{\chi^2}\xspace}

\newcommand{\Mstar}{\ensuremath{M_{\star}}\xspace}
\newcommand{\Rstar}{\ensuremath{R_{\star}}\xspace} 
\newcommand{\Lstar}{\ensuremath{L_{\star}}\xspace} 
\newcommand{\fe}{\ensuremath{\mathrm{[Fe/H]}}\xspace}
\newcommand{\teff}{\ensuremath{T_{\mathrm{eff}}}\xspace}  
\newcommand{\logg}{\ensuremath{\log g}\xspace} 
\newcommand{\vsini}{\ensuremath{v \sin i}\xspace} 
\newcommand{\Fbol}{\ensuremath{F_{\mathrm{bol}}}\xspace}
\newcommand{\logage}{\ensuremath{\log_{10}{(\mathrm{age})}}\xspace}
\newcommand{\plx}{\ensuremath{\pi_\star}\xspace}

\newcommand{\kms}{km s$^{-1}$\xspace}

\newcommand{\Rsun}{\ensuremath{R_{\odot}}\xspace }
\newcommand{\Msun}{\ensuremath{M_{\odot}}\xspace}

\newcommand{\angstrom}{\AA\xspace}

\newcommand{\libnum}{404\xspace}
\newcommand{\kdwarfnum}{23\xspace}
\newcommand{\mannnum}{56\xspace}
\newcommand{\brunttnum}{55\xspace}
\newcommand{\brewernum}{177\xspace}
\newcommand{\vonbraunnum}{95\xspace}
\newcommand{\libtefflo}{3000\xspace}
\newcommand{\libteffhi}{7000\xspace}
\newcommand{\libRstarlo}{0.1\xspace}
\newcommand{\libRstarhi}{16\xspace}

\newcommand{\nwav}{73788\xspace}

\newcommand{\sigteff}{100 K\xspace}
\newcommand{\sigRstar}{15\%\xspace}
\newcommand{\sigfe}{0.09 dex\xspace}

\newcommand{\sigteffcool}{70 K\xspace}
\newcommand{\sigRstarcool}{10\%\xspace}
\newcommand{\sigfecool}{0.12 dex\xspace}

\begin{document}

\title{Precision Stellar Characterization of FGKM Stars using an Empirical Spectral Library}

\author{Samuel W. Yee\altaffilmark{1,2}}
\author{Erik A. Petigura\altaffilmark{1,3}}
\and
\author{Kaspar von Braun\altaffilmark{4}}

\altaffiltext{1}{California Institute of Technology}
\altaffiltext{2}{syee@caltech.edu}
\altaffiltext{3}{Hubble Fellow}
\altaffiltext{4}{Lowell Observatory, 1400 W. Mars Hill Rd., Flagstaff, AZ 86001}

\begin{abstract}
Classification of stars, by comparing their optical spectra to a few dozen spectral standards, has been a workhorse of observational astronomy for more than a century. Here, we extend this technique by compiling a library of optical spectra of \libnum touchstone stars observed with Keck/HIRES by the California Planet Search. The spectra are high-resolution ($R\approx60000$), high signal-to-noise (SNR $\approx$ 150/pixel), and registered onto a common wavelength scale. The library stars have properties derived from interferometry, asteroseismology, LTE spectral synthesis, and spectrophotometry. To address a lack of well-characterized late K-dwarfs in the literature, we measure stellar radii and temperatures for \kdwarfnum nearby K-dwarfs, using SED modeling and \Gaia parallaxes. This library represents a uniform dataset spanning the spectral types $\sim$M5--F1 ($\teff\approx\libtefflo-\libteffhi$~K, $\Rstar\approx\libRstarlo-\libRstarhi$~\Rsun). We also present ``Empirical SpecMatch'' (\SpecMatch), a tool for parameterizing unknown spectra by comparing them against our spectral library. For FGKM stars, \SpecMatch achieves accuracies of \sigteff in effective temperature (\teff), \sigRstar in stellar radius (\Rstar), and \sigfe in metallicity (\fe). Because the code relies on empirical spectra it performs particularly well for stars $\sim$K4 and later which are challenging to model with existing spectral synthesizers, reaching accuracies of \sigteffcool in \teff, \sigRstarcool in \Rstar, and \sigfecool in \fe. We also validate the performance of \SpecMatch, finding it to be robust at lower spectral resolution and SNR, enabling the characterization of faint late-type stars. Both the library and stellar characterization code are publicly available.
\end{abstract}


\section{Introduction}
Measuring the physical properties of stars is a long-standing and important problem in astronomy. The masses, radii, and temperatures of stars are benchmarks against which we test models of stellar structure and evolution. The abundances of iron and other elements in stellar populations help trace the nucleosynthetic enrichment history of the Milky Way. Recently, the study of extrasolar planets has placed new demands on {\em precision} stellar characterization. The extent to which observational methods like radial velocities and transit photometry can reveal a planet's mass and radius is limited by the uncertainty in stellar mass and radius, respectively.

Eclipsing binaries typically offer the most precise and least model-dependent measurements of stellar mass (\Mstar) and radius (\Rstar), relying primarily on Newtonian mechanics and geometry. Analysis of eclipsing binary light curves and radial velocities can achieve measurements of \Mstar and \Rstar good to $\lesssim3\%$. After incorporating broadband photometry and parallaxes, eclipsing binaries can yield effective temperatures (\teff) good to $\sim2\%$ \citep{Torres10}.

Measurements of such stellar properties for isolated field stars are more challenging and rely more heavily on models of stellar interiors and atmospheres. Recently, precision space-based photometry from the {\em Kepler} and {\em CoRoT} missions \citep{Borucki10,Auvergne09} have enabled the detection of stellar acoustic modes which achieve \Mstar and \Rstar good to a few \% (e.g. \citealp{Bruntt12}, \citealp{Huber13}). The amplitude of these modes grows with the size of the star and have been detected in stars that are roughly solar-size or larger. For smaller stars, the acoustic oscillation amplitudes are smaller, making their detection increasingly difficult against photometric noise due to photon Poisson statistics, surface granulation, and other sources of stellar activity. While asteroseismology has provided some excellent stellar benchmarks, it is not applicable for all stellar types. This illustrates a recurring challenge in the field of precision stellar astrophysics: no single technique is uniformly effective across the Hertzsprung-Russell (HR) diagram. 

Recently, optical and infrared interferometry have been used to directly measure the angular size of stars (e.g. \citealp{VonBraun14}). When combined with parallax measurements and an observed spectral energy distribution, this technique provides \teff and \Rstar that are almost purely empirical. Interferometry requires stars that are nearby and bright and thus currently applicable to only a limited number. As of 2016, only about $\sim100$ main sequence stars and $\sim200$ giants have precision interferometric measurements. Stars cooler than $\sim5000$~K are underrepresented in this interferometric sample due to their faintness. Finally, obtaining \Mstar relies on a spectroscopic measurement of \logg coupled with the directly determined \Rstar or by using stellar structure models with \teff, \Rstar, and spectroscopic measurements of the star's metallicity as input.

Another common technique for stellar characterization relies on detailed modeling of a star's spectrum. This involves constructing a model stellar atmosphere and modeling the radiative transfer of photons as they emerge from the photosphere on their way toward Earth. The effective temperature, surface gravity (\logg), metallicity (\fe), abundance of other species, and projected rotational velocity (\vsini) are adjusted until the simulated distribution of photons matches the observed spectrum. Two commonly-used spectral synthesis codes are {\tt MOOG} \citep{Sneden73} and {\tt SME} \citep{Valenti96,Valenti05,Brewer15}. One challenge facing spectral synthesis is that model stellar atmospheres grow more uncertain as they diverge from solar. Consequently, such codes may accurately reproduce the observed stellar spectrum, but do so with parameters that may be different from the true stellar properties. Spectral synthesis techniques begin to suffer dramatically for stars having spectral type K4 and later ($\teff \lesssim 4700$~K), due to the large number of atomic and molecular lines combined with missing or inaccurate values in the input line lists and associated quantum mechanical properties.

With this background in mind, we present a new method, ``Empirical SpecMatch'' (\SpecMatch), for precision stellar characterization based on the direct comparison of observed optical spectra to a dense library of touchstone stars with well-measured properties. A chief goal of \SpecMatch is accuracy across the HR diagram with a specific focus on stars of mid-K spectral type and later, due to the uncertainties with associated with synthetic spectral techniques.

\SpecMatch builds off a long history of stellar classification in astronomy using optical spectra. Founding work took place at the Harvard College Observatory (c. 1885--1925) with the visionary and heroic efforts of A.~J.~Cannon, A.~P.~Draper, and E.~C.~Pickering which culminated in the HD catalog and the OBAFGKM classification sequence. A significant refinement by \cite{Morgan43} included luminosity classes as a essential second dimension in the sequence \citep{Gray09}.

\SpecMatch extends this methodology by considering a third dimension, metallicity, and by returning numerical measurements of \teff, \Rstar, and \fe of stellar parameters as opposed to categorical spectral classifications. This approach is enabled by the large and homogeneous sample of high resolution, high signal-to-noise spectra gathered by the California Planet Search as part of its radial velocity programs to study extrasolar planets and by recent work to compile catalogs of touchstone stars with well-measured properties.

\SpecMatch consists of two major parts: a spectral library and an algorithm for parameter measurement. In Section~\ref{sec:library}, we describe the construction of the spectral library of stars having properties determined by asteroseismology, interferometry, spectroscopy, and spectrophotometry. To address the scarcity of suitable mid- to late-K dwarfs from the literature, we present new precision measurements of \kdwarfnum K-dwarfs incorporating spectroscopy, SED modeling, and parallax measurements. This library of spectra and stellar parameters are publicly available for download as a monolithic memory-efficient Hierarchical Data Format (HDF5) file\footnote{\url{https://zenodo.org/record/168084/files/library.h5}} or as individual Flexible Image Transport System (FITS) files for each spectrum.\footnote{\url{http://web.gps.caltech.edu/\~syee/hires_spectra/}}

In Section~\ref{sec:specmatchalg}, we describe the \SpecMatch algorithm, which compares an unlabelled target spectrum with the spectral library. The code, as well as easy-to-use command-line scripts, can also be obtained online.\footnote{\url{https://github.com/samuelyeewl/specmatch-emp}} In Section~\ref{sec:performance}, we assess the accuracy of \SpecMatch by performing an internal cross-validation analysis. We verified that for stars ranging from $\teff\approx\libtefflo- \libteffhi$~K, \SpecMatch yields \teff, \Rstar, and \fe good to \sigteff, \sigRstar, and \sigfe respectively. We also investigate the performance of the algorithm at decreased SNR and spectral resolution by degrading the target spectra. We find that \SpecMatch remains robust even at SNR as low as 10/pixel, as well as spectral resolutions down to about $R = 20000$.

\section{Library} \label{sec:library}
\begin{figure}
	\epsscale{1.2}
	\plotone{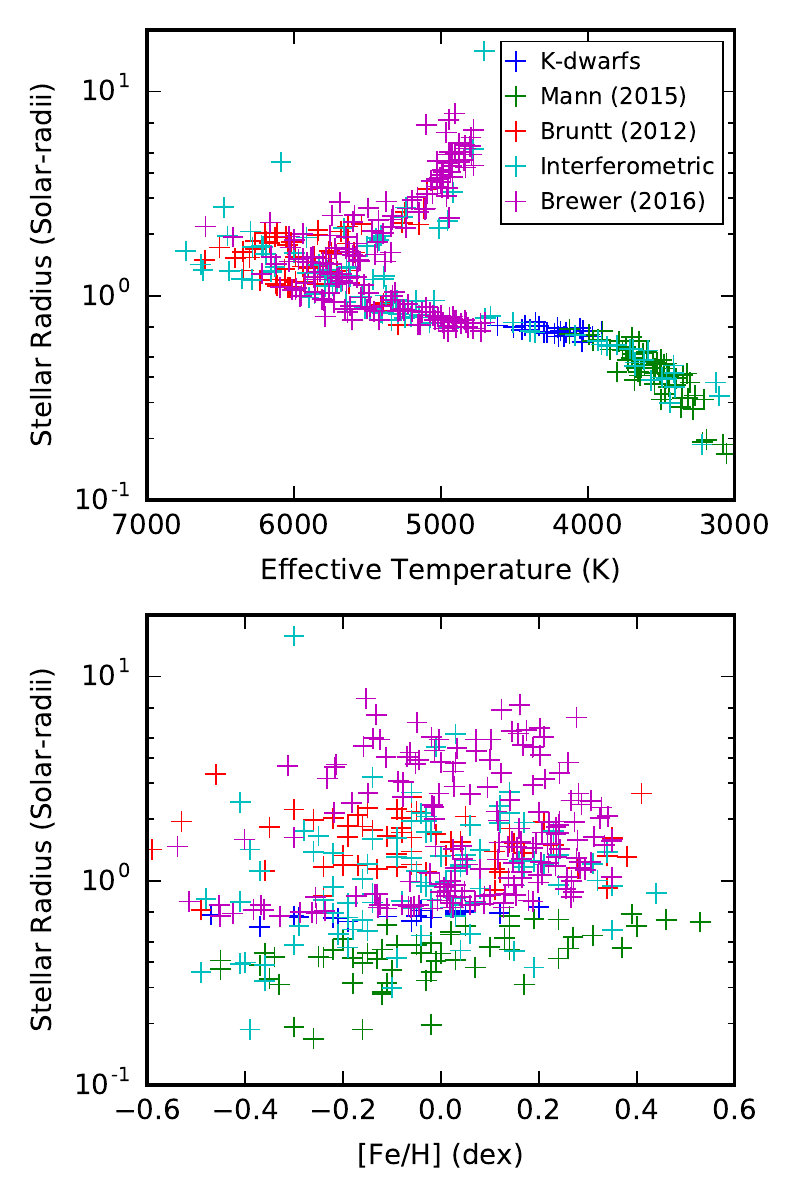}
	\caption{Distribution of SpecMatch-Empirical library stars. The library covers the parameter space of $\teff \approx 3000-7000$~K, $\Rstar \approx 0.16-16$~\Rsun, $\fe \approx -0.5-+0.5$~dex. We have excluded stars having measured $\vsini > 8.0$~\kms as their spectra retain insufficient information for the matching process. \label{fig:library}}
\end{figure}

A crucial component of the \SpecMatch is the spectral library of \libnum stars, each with well-determined stellar parameters (\Mstar, \Rstar, \teff, \fe). As no single technique yields the best parameters across the HR diagram, we have compiled the parameters for the stars in our library from a variety of different sources. By construction, our library contains stars that span a large region of the HR diagram ($\teff\approx\libtefflo-\libteffhi$~K, $\Rstar\approx\libRstarlo-\libRstarhi$~\Rsun). The domain of \teff, \Rstar, and \fe is shown in Figure~\ref{fig:library} along with the provenance of the library parameters. We describe in brief the source catalogs for these parameters in Section~\ref{sec:library_params}. We augment this sample with new K-dwarf library stars in Section~\ref{ssec:kdwarfs}. In Section~\ref{ssec:isochrones}, we describe our conversion of the measured stellar properties into a homogeneous suite of \teff, \logg, \fe, \Mstar, \Rstar, and age measurements for each star. The details of the spectral dataset are described in Section~\ref{ssec:library_spectra}.

\subsection{Library Parameters} \label{sec:library_params}

\subsubsection{Interferometric sample} \label{sec:sample_interferometric}
The library includes \vonbraunnum stars having interferometrically-determined radii with quoted uncertainties in radius of $< 5\%$ compiled by \cite{VonBraun14}.%
\footnote{
The \cite{VonBraun14} compilation included results from \cite{Kervella03}, \cite{Baines08}, \cite{Baines09}, \cite{VanBelle09}, \cite{VonBraun11a}, \cite{VonBraun11b}, \cite{VonBraun12}, \cite{Boyajian13}, and \cite{Henry13}.
}
This set of interferometric stars is the only sample included in our library to span the entire HR diagram, as the other techniques face limitations in different regions of the parameter space. However, the relatively small number of stars with interferometric measurements necessitated the inclusion of stars from other sources.

\subsubsection{Mann et al. 2015} \label{sec:sample_mann}
\cite{Mann15} measured fundamental properties of 183 late K and M dwarfs by performing an absolute flux calibration on optical and NIR spectra using literature photometry. The photometric measurements were calibrated with the updated filter profiles from \cite{Mann2015b}. The bolometric flux (\Fbol) was calculated by integrating over the flux-calibrated spectra, while \teff was obtained by comparing the spectra with PHOENIX stellar atmosphere models. \Fbol and \teff were then combined distance, $d$, measured by trigonometric parallax, to obtain the physical radius of the star (\Rstar) using the Stefan-Boltzmann law,
\[
\Lstar = 4 \pi d^2 \Fbol = 4 \pi \Rstar^2 \sigma \teff^4.
\]
The analysis used empirical relations between spectral feature widths and \fe from \cite{Mann2013} and \cite{Mann2014} to obtain the stellar metallicities. \Mstar was determined using empirical mass-luminosity relations from \cite{Delfosse2000}. Both of these empirical relationships have been calibrated for M dwarfs. Median uncertainties for the stellar properties were 60~K in \teff, 3.8\% in \Rstar, 0.08~dex in \fe and 10\% in \Mstar. Our library includes \mannnum of the stars from this sample, for which we had suitable HIRES spectra. These comprise the majority of the M dwarfs in our library.

\subsubsection{Brewer et al. 2016} 
\label{sec:sample_brewer}
\cite{Brewer16} conducted a detailed spectroscopic analysis of 1617 dwarf and subgiant stars with \teff~=~4700--6800~K observed by Keck/HIRES. \cite{Brewer16} modeled the spectra with {\tt SME} \citep{Valenti96} which has been recently updated to included many more lines \citep{Brewer15}. We chose \brewernum of the stars from this analysis for our library, excluding stars with $\vsini > 8 $~\kms, SNR < 120 per pixel, and $\fe < -1.0$~dex. We adopted the following uncertainties for the stellar parameters: 60 K in \teff, 0.05 dex in \logg, and 0.05 dex in \fe based on the comparisons with other independent techniques performed in \cite{Brewer15} and \cite{Brewer16}.

\subsubsection{Bruntt et al. 2012} 
\label{sec:sample_asteroseismic}

\cite{Bruntt12} examined 93 solar-type stars observed by \Kepler, combining asteroseismology and spectroscopy to measure \teff, \logg, and \fe. The quoted uncertainties on these properties were 0.03 dex in \logg, 60 K in \teff, and 0.06 dex in \fe. We included \brunttnum stars from this sample with HIRES spectra in our library.

\subsection{K Dwarfs}
\label{ssec:kdwarfs}
Finding well-characterized mid to late K-dwarfs (K4-K7; $\teff \approx 4700$--$4000$~K) proved to be particularly challenging. Detailed spectral synthesis codes such as {\tt SME} and {\tt MOOG} are challenged by the proliferation of deep metallic lines and the onset of MgH and TiO bands \citep{Gray09}. The empirical relations used in \cite{Mann15} have not been calibrated for spectral types earlier than K7 ($\teff\lesssim4100$~K). While measurements based on CHARA interferometry exist for a few such stars, coverage is sparse due to the small number of stars that are both bright enough at optical wavelengths and close enough for precise radius measurements.

Due to the paucity of touchstone stars in the literature, we derived \teff and \Rstar for \kdwarfnum K4--K7 dwarfs using broadband photometry and constraints from parallax, \plx. We drew stars from a sample of 110 included in a survey of planets around late K-dwarfs \citep{Gaidos13}. After a search of literature photometry, we restricted this sample to the \kdwarfnum stars having at least one literature measurement in the {\em UBVRIJHK} bands, which provide good coverage across the stellar Spectral Energy Distribution (SED).

We derived \teff using an empirical $V-K$ color-temperature relationship from \cite{Boyajian13}
\[
\label{eq:colortemp}
\teff = a_0 + a_1 (V - K) + a_2 (V - K)^2 + a_3 (V - K)^3
\]
where
\begin{align*}
&a_0 = 8649 \pm 28~\mathrm{K}\\
&a_1 = -2503 \pm 35~\mathrm{K}\, \mathrm{mag}^{-1}\\
&a_2 = 442 \pm 12~\mathrm{K}\, \mathrm{mag}^{-2}\\
&a_3 = -31.7 \pm 1.5~\mathrm{K}\, \mathrm{mag}^{-3}
\end{align*}
The coefficients $a_i$ are found in Table~8 of \cite{Boyajian13} and were calibrated to 125 FGK stars with literature photometry and effective temperatures. Using this relation with existing literature photometry for our sample of K dwarfs, we were able to find \teff to a typical uncertainty of $\sim 5\%$.

Next, we measured \Fbol by fitting the broadband photometry with a stellar SED and integrating over wavelength. Our procedure for the SED fitting follows \citet{VonBraun14}: we used the library of spectral templates from \citet{Pickles1998} and fit them to literature photometry data via \chisq-minimization. We linearly interpolated between the published spectral templates to the nearest half spectral type to improve \chisq fitting. For the literature photometry data, we make use of the modified filter profiles from \citet{Mann2015b}. Reddening is set to zero for all targets due to their small distances (< 26 pc). After the template is scaled to minimize the reduced $\chi^2$, bolometric flux is measured by integrating over wavelength.

Twenty one stars from our sample have updated parallaxes from new Gaia data, as part of the Tycho-Gaia Astrometric Solution (TGAS; \citealt{Gaia16a,Lindegren2016,Michalik2015}). These stars have uncertainties in parallax, $\sigma(\plx) \approx 0.3$~mas. For the remaining two stars (GJ 1172 and HD 85488), we used the parallax values calculated in the van Leeuwen reduction of {\em Hipparcos} data with $\sigma(\plx)\approx1.6$~mas \citep{Perryman97,VanLeeuwen07}. Combining \teff, \Fbol, \plx and the Stefan-Boltzmann law, we calculated \Rstar. The median uncertainty in \Rstar over the entire sample was $7.4\%$, dominated primarily by the uncertainties in \teff from the color-temperature relation. 

Finally, we adopted the metallicity values found in \cite{Gaidos13}, which were derived using \texttt{SME} \citep{Valenti96}. While traditional spectroscopic techniques begin to suffer for $\teff \lesssim 4700$~K, they are the only source of metalicity measurements for this sample of K dwarfs. More accurate metallicities could be derived by observing K-dwarfs with G-dwarf companions for which metallicity can be measured more accurately from spectral synthesis. Such a study is beyond the scope of this work. We adopt 0.1~dex as the uncertainty on \fe and note that such errors are likely systematic (rather than statistical) in nature. The final set of parameters used for these K dwarfs are listed in Table~\ref{table:kdwarfs}.

\onecolumngrid
\begin{deluxetable*}{llrrrcrrrc}
\tablecaption{Properties of \kdwarfnum K-Dwarfs\label{table:kdwarfs}}
\tabletypesize{\scriptsize}
\tablecolumns{9}
\tablewidth{0pt}
\tablehead{
	\colhead{Name} &
	\colhead{HIP} & 
	\colhead{$V$} & 
	\colhead{$K$} &
	\colhead{\teff} & 
	\colhead{\Fbol} &
	\colhead{\plx} & 
	\colhead{\Rstar} &
    \colhead{\fe} &
    \colhead{Notes} \\
	\colhead{} &
	\colhead{} & 
	\colhead{mag} & 
	\colhead{mag} &
	\colhead{K} & 
	\colhead{$10^{-8} \mathrm{erg}^{-1} \mathrm{cm}^{-2}$} &
	\colhead{mas} & 
	\colhead{\Rsun} &
    \colhead{dex} &
    \colhead{}
}
\startdata
GJ 1008.0     & 1532     & 9.92 $\pm$ 0.04     & 6.58 $\pm$ 0.03     & 4039 $\pm$ 241     & 0.66 $\pm$ 0.01     & 49.3 $\pm$ 0.4     & 0.59 $\pm$ 0.06     & -0.37 $\pm$ 0.10     & 1 \\
GJ 1044     & 10337     & 9.86 $\pm$ 0.04     & 6.54 $\pm$ 0.03     & 4051 $\pm$ 239     & 0.68 $\pm$ 0.01     & 42.6 $\pm$ 0.9     & 0.69 $\pm$ 0.07     & 0.12 $\pm$ 0.10     & 1 \\
GJ 1127     & 47201     & 9.45 $\pm$ 0.03     & 6.37 $\pm$ 0.03     & 4207 $\pm$ 212     & 0.86 $\pm$ 0.01     & 44.5 $\pm$ 0.2     & 0.70 $\pm$ 0.05     & 0.03 $\pm$ 0.10     & 1 \\
GJ 1172     & 66222     & 9.95 $\pm$ 0.04     & 6.38 $\pm$ 0.03     & 3904 $\pm$ 266     & 0.71 $\pm$ 0.01     & 48.8 $\pm$ 1.6     & 0.67 $\pm$ 0.09     & -0.11 $\pm$ 0.10     & 2 \\
GJ 3072     & 4845     & 10.03 $\pm$ 0.04     & 6.57 $\pm$ 0.06     & 3967 $\pm$ 301     & 0.62 $\pm$ 0.01     & 46.7 $\pm$ 0.4     & 0.63 $\pm$ 0.07     & -0.19 $\pm$ 0.10     & 1 \\
GJ 3494     & 40910     & 9.76 $\pm$ 0.04     & 6.61 $\pm$ 0.03     & 4159 $\pm$ 228     & 0.68 $\pm$ 0.01     & 44.3 $\pm$ 0.3     & 0.63 $\pm$ 0.05     & -0.06 $\pm$ 0.10     & 1 \\
GJ 9093     & 12493     & 9.52 $\pm$ 0.03     & 6.54 $\pm$ 0.03     & 4276 $\pm$ 193     & 0.74 $\pm$ 0.01     & 41.9 $\pm$ 0.3     & 0.66 $\pm$ 0.05     & -0.29 $\pm$ 0.10     & 1 \\
GJ 9144     & 19165     & 9.69 $\pm$ 0.03     & 6.74 $\pm$ 0.03     & 4298 $\pm$ 200     & 0.64 $\pm$ 0.01     & 38.7 $\pm$ 0.2     & 0.66 $\pm$ 0.05     & -0.22 $\pm$ 0.10     & 1 \\
HD 178126     & 93871     & 9.22 $\pm$ 0.02     & 6.47 $\pm$ 0.03     & 4449 $\pm$ 177     & 0.87 $\pm$ 0.01     & 40.8 $\pm$ 0.3     & 0.68 $\pm$ 0.04     & -0.47 $\pm$ 0.10     & 1 \\
HD 200779     & 104092     & 8.30 $\pm$ 0.02     & 5.33 $\pm$ 0.04     & 4283 $\pm$ 211     & 2.47 $\pm$ 0.01     & 66.6 $\pm$ 0.2     & 0.76 $\pm$ 0.06     & 0.07 $\pm$ 0.10     & 1 \\
HD 20280     & 15095     & 9.16 $\pm$ 0.02     & 6.11 $\pm$ 0.03     & 4227 $\pm$ 199     & 1.14 $\pm$ 0.01     & 54.0 $\pm$ 0.3     & 0.65 $\pm$ 0.05     & -0.21 $\pm$ 0.10     & 1 \\
HD 203040     & 105341     & 9.10 $\pm$ 0.04     & 5.75 $\pm$ 0.03     & 4033 $\pm$ 242     & 1.38 $\pm$ 0.01     & 63.2 $\pm$ 0.5     & 0.67 $\pm$ 0.06     & -0.05 $\pm$ 0.10     & 1 \\
HD 218294     & 114156     & 9.62 $\pm$ 0.02     & 6.45 $\pm$ 0.03     & 4146 $\pm$ 209     & 0.75 $\pm$ 0.01     & 45.1 $\pm$ 0.3     & 0.66 $\pm$ 0.05     & -0.02 $\pm$ 0.10     & 1 \\
HD 224607     & 118261     & 8.70 $\pm$ 0.05     & 6.15 $\pm$ 0.03     & 4615 $\pm$ 214     & 1.29 $\pm$ 0.01     & 44.0 $\pm$ 0.2     & 0.71 $\pm$ 0.05     & -0.04 $\pm$ 0.10     & 1 \\
HD 35171     & 25220     & 7.94 $\pm$ 0.03     & 5.26 $\pm$ 0.03     & 4505 $\pm$ 179     & 2.74 $\pm$ 0.02     & 68.5 $\pm$ 0.3     & 0.70 $\pm$ 0.04     & 0.05 $\pm$ 0.10     & 1 \\
HD 355784     & 97051     & 9.96 $\pm$ 0.05     & 6.87 $\pm$ 0.03     & 4200 $\pm$ 248     & 0.53 $\pm$ 0.01     & 38.5 $\pm$ 0.3     & 0.63 $\pm$ 0.06     & -0.19 $\pm$ 0.10     & 1 \\
HD 59582     & 36551     & 8.98 $\pm$ 0.03     & 6.16 $\pm$ 0.03     & 4395 $\pm$ 193     & 1.13 $\pm$ 0.01     & 48.5 $\pm$ 0.3     & 0.67 $\pm$ 0.05     & -0.30 $\pm$ 0.10     & 1 \\
HD 68834     & 40375     & 8.82 $\pm$ 0.04     & 5.92 $\pm$ 0.03     & 4334 $\pm$ 209     & 1.39 $\pm$ 0.01     & 52.0 $\pm$ 0.3     & 0.71 $\pm$ 0.05     & 0.03 $\pm$ 0.10     & 1 \\
HD 7279     & 5663     & 9.56 $\pm$ 0.04     & 6.45 $\pm$ 0.03     & 4186 $\pm$ 222     & 0.77 $\pm$ 0.01     & 44.0 $\pm$ 0.3     & 0.67 $\pm$ 0.06     & -0.05 $\pm$ 0.10     & 1 \\
HD 80632     & 45839     & 9.10 $\pm$ 0.04     & 6.32 $\pm$ 0.03     & 4426 $\pm$ 198     & 0.99 $\pm$ 0.01     & 41.8 $\pm$ 0.2     & 0.71 $\pm$ 0.05     & 0.05 $\pm$ 0.10     & 1 \\
HD 85488     & 48411     & 8.85 $\pm$ 0.03     & 5.98 $\pm$ 0.03     & 4357 $\pm$ 199     & 1.30 $\pm$ 0.01     & 47.7 $\pm$ 1.5     & 0.74 $\pm$ 0.07     & 0.20 $\pm$ 0.10     & 2 \\
HD 97214     & 54651     & 9.23 $\pm$ 0.03     & 6.36 $\pm$ 0.03     & 4357 $\pm$ 197     & 0.87 $\pm$ 0.01     & 50.5 $\pm$ 0.4     & 0.58 $\pm$ 0.04     & -0.89 $\pm$ 0.10     & 1 \\
HD 97503     & 54810     & 8.70 $\pm$ 0.02     & 5.88 $\pm$ 0.03     & 4395 $\pm$ 185     & 1.46 $\pm$ 0.01     & 54.2 $\pm$ 0.3     & 0.68 $\pm$ 0.04     & 0.03 $\pm$ 0.10     & 1 \\

\enddata
\tablecomments{Parameters of newly-characterized K4-K7 dwarfs determined by combining an empirical color-temperature relation, SED-fitting, and stellar parallaxes.}
\tablenotetext{1}{Parallax from the Tycho-Gaia Astrometric Solution (TGAS; \citealt{Gaia16a,Lindegren2016,Michalik2015}).}
\tablenotetext{2}{Parallax from Hipparcos.}
\end{deluxetable*}

\subsection{Isochrone Analysis}
\label{ssec:isochrones}
The catalogs of stars presented in Section~\ref{sec:library_params} often do not list a complete set of \teff, \Rstar, \Mstar, \logg, and \fe. For example, the interferometric sample (Section~\ref{sec:sample_interferometric}) does not include measurements of \Mstar. To obtain a homogeneous suite of parameters, we combine the known properties with the Dartmouth models \citep{Dotter2008}. We use the \isochrones software package \citep{isochrones}, which performs an MCMC fit to estimate the remaining stellar parameters from the stellar model grid. To account for systematic errors in the Dartmouth models, we adopt the 5th and 95th percentiles of the final MCMC distribution as the parameter uncertainties.

\subsection{Optical Spectra} 
\label{ssec:library_spectra}
The spectra were observed with the High Resolution Echelle Spectrometer (HIRES; \citealt{Vogt94}) at the Keck-I 10-m telescope as part of the various programs by the California Planet Search to study extrasolar planets. For information about CPS and its goals, see \cite{Howard10}. The spectra have a resolution, $R \approx$ 50000--60000. We selected stars with spectra having SNR of at least 40 per pixel, with most ($\sim80\%$) having signal-to-noise ratio (SNR) of >100 per HIRES pixel on blaze near 550~nm. 

The extracted spectra are from the middle chip of HIRES and contain 16 orders with 4021 intensity measurements per order. Each order is imprinted with the HIRES blaze function which we remove by dividing the observed spectrum with a calibration spectrum constructed from several rapidly rotating B-stars. Due to different line of sight velocities, the library spectra are shifted with respect to one another in pixel-coordinates. It is convenient to adopt a common wavelength scale and register each spectrum to this scale. We resampled our spectra onto a new wavelength scale ($\lambda$ = 4990--6410 \angstrom) that is uniform in $\Delta \log \lambda$. We then removed the shifts due to the stars' individual line of sight velocities by registering our spectra against the National Solar Observatory (NSO) solar spectrum \citep{Kurucz84} according to the procedure described in Section~\ref{ssec:shifting}. The final library data set thus consists of a $\libnum \times \nwav$ array of normalized spectra on the rest wavelength scale, together with associated stellar parameters.

\section{SpecMatch} 
\label{sec:specmatchalg}
Here, we describe the \SpecMatch algorithm. Given an unknown spectrum, \SpecMatch first shifts it onto the library wavelength scale to correct for their individual radial velocity (Section~\ref{ssec:shifting}). We then identify the most similar library spectra (Section~\ref{ssec:matching}) and interpolate between them (Section~\ref{ssec:lincomb}) to arrive at a final set of derived parameters for the target star.

We report \teff, \Rstar, and \fe in contrast to most spectral analysis codes, which report \teff, \logg, and \fe. While \Rstar is closely related to \logg, for the library stars K4 and later, \Rstar is the directly observed quantity. We could, in principle, derive \logg values by appealing to stellar models, but such models have known systematic errors that are largest for cool stars. For example, in their study of M-dwarfs, \cite{Mann15} found that the Dartmouth models systematically over-predict stellar radii by $\approx 5\%$. Therefore, to mitgate systematic errors in the derived stellar radii of late-type stars, where \SpecMatch performs best, we choose to parametrize stars in terms of \teff, \Rstar, and \fe.

\subsection{Spectral Registration} 
\label{ssec:shifting}
We begin by resampling each of the 16 orders of the HIRES spectrum onto the library wavelength scale. Because the library wavelength scale is uniform in $\Delta \log \lambda$, fixed velocity shifts between the spectra correspond a uniform displacement in the current logarithmic wavelength sampling. We also mask out the telluric lines due to atmospheric O$_2$ absorption in the $\lambda = 6270-6310$~\angstrom wavelength region, as these lines are fixed in the observatory frame, not in the stellar rest frame. 

\begin{figure}
	\epsscale{1.2}
	\plotone{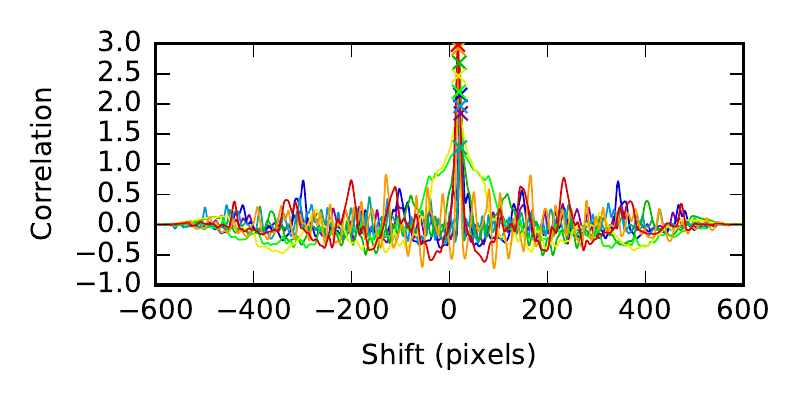}
	\epsscale{1.2}
	\plotone{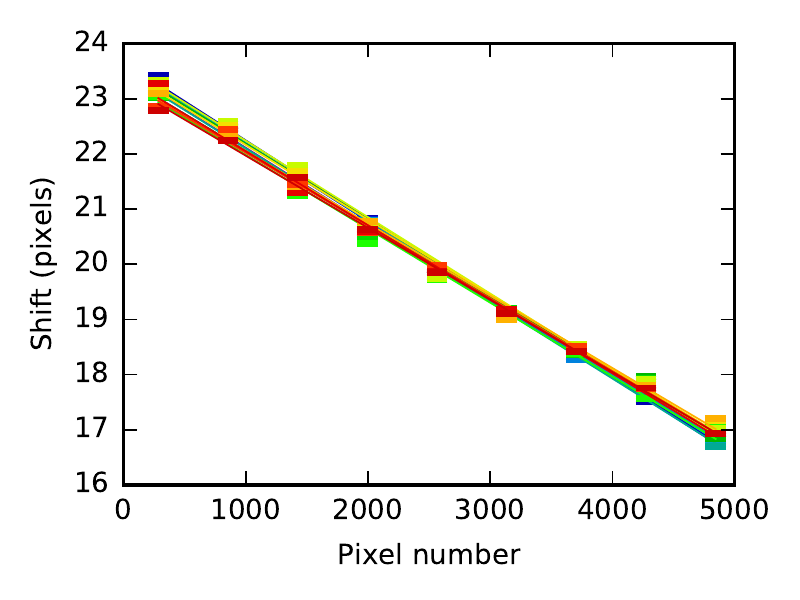}
	\caption{Registration of a solar-type spectrum using cross-correlation (see \autoref{ssec:shifting}). Top: Cross-correlation of a single order of the spectrum of HD190406 (spectral type G0V) with corresponding regions of the solar spectrum. Each line corresponds to a particular segment. The maximum correlation for each segment is indicated with a cross. Bottom: The shifts calculated for the various segments of each order are fit with a first order polynomial, which is then applied to shift the corresponding pixels. Outliers result from poor cross-correlation results in that region of spectrum and are clipped before the fit is performed.\label{fig:corr_lags}}
\end{figure}

\begin{figure*}
    \epsscale{1.15}
	\plotone{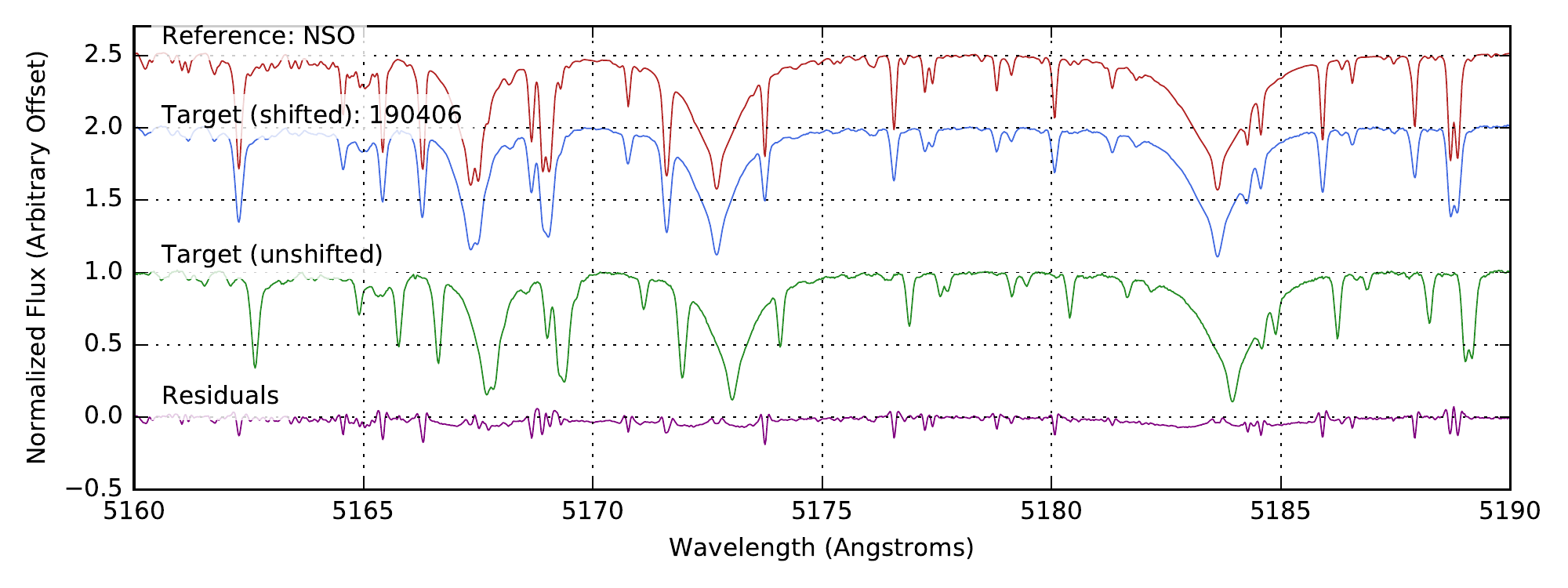}
	\caption{Solar reference spectrum and spectrum of HD190406 (spectral type G0V) before and after registration (see \autoref{ssec:shifting}). \label{fig:shifted_spec}}
\end{figure*}

Each masked order is then divided into segments of length $\sim 700$ pixels, which are then cross-correlated with the reference spectrum. This operation is performed in Fourier space and low frequency Fourier modes are filtered out such that only spectral lines, and not long-baseline differences in continuum normalization, contribute to the cross-correlation. For each segment, we find the ideal pixel shift by fitting a parabola to the cross-correlation peak. We fit a line through these shifts to obtain a linear relationship between the pixel shift and pixel number across each order (Figure~\ref{fig:corr_lags}). As an example, we show the observed and registered spectrum of HD190406 in Figure~\ref{fig:shifted_spec}.

\begin{figure*}
	\epsscale{1.15}
	\plottwo{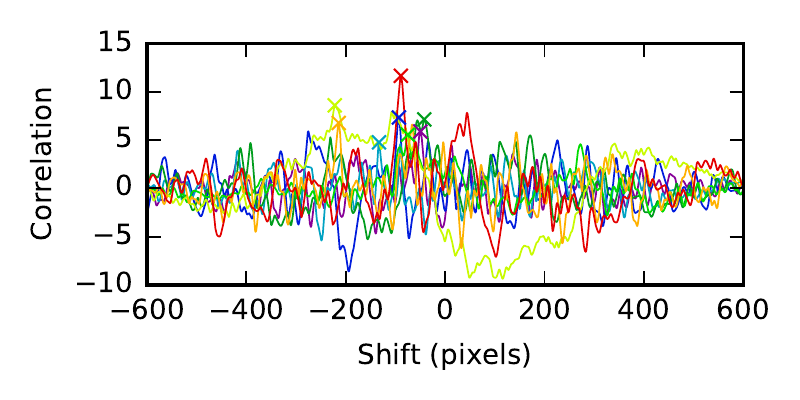}{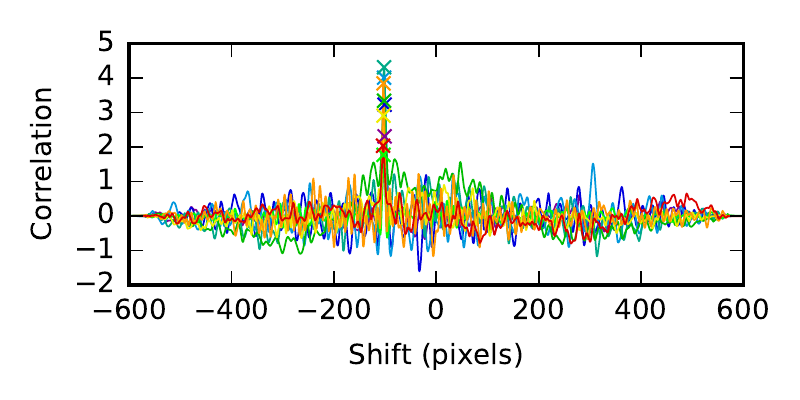}
	\caption{Cross-correlation of the different segments in a single order of the spectrum of Barnard's star (GL669, spectral type M4V), against the solar spectrum (left) and an M dwarf spectrum (right). Due to the dissimilarity in spectra, the cross-correlation with the solar spectrum gives multiple peaks which do not correspond to the true shift, whereas the previously-shifted M dwarf spectrum serves as a better reference (see \autoref{ssec:shifting}). \label{fig:bootstrap_corr}}
\end{figure*}

As a matter of convenience, all the library spectra are shifted onto a single wavelength scale, chosen to be that of the NSO solar spectrum. However, for stars with spectral types unlike the Sun, performing the cross-correlation against the NSO spectrum gives poor results, with multiple spurious peaks which do not correspond to the true velocity shift (see top panel of Figure~\ref{fig:bootstrap_corr}).

\begin{figure*}
    \epsscale{1.15}
	\plotone{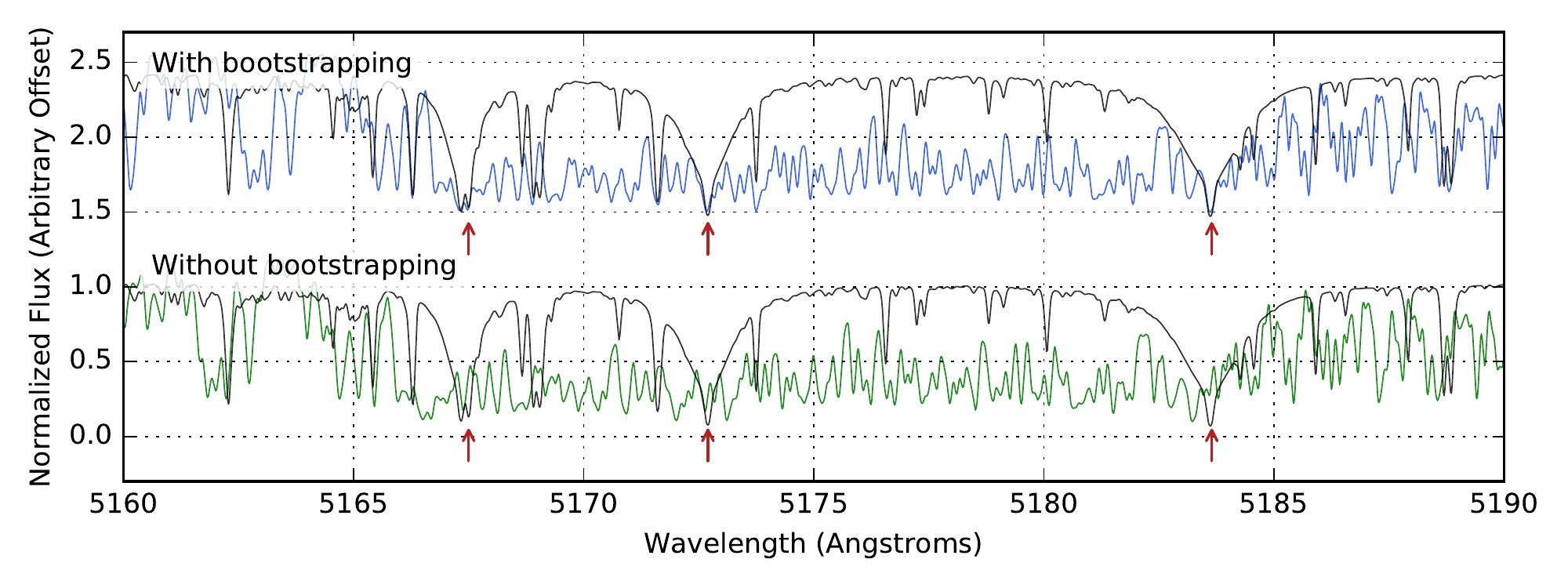}
	\caption{Spectrum of Barnard's star (GL699, spectral type M4V) when shifted against the solar spectrum (bottom) and against a previously shifted template M dwarf spectrum (top). The solar spectrum, which was chosen to be the library rest wavelength scale, is overlaid in gray. Red arrows indicate specific spectral features which are properly aligned only when the target spectrum is shifted against the M dwarf reference. Using the bootstrapping approach described in \autoref{ssec:shifting} to derive a ladder of wavelength standards of different spectral types, we achieve superior spectral registration. \label{fig:bootstrap_spectrum}}
\end{figure*}

To address this, we use a bootstrapping approach in which four additional spectra belonging to stars of different spectral types were shifted successively onto the NSO wavelength scale, forming a ladder of template spectra. These additional spectra were chosen from our highest signal-to-noise spectra (SNR > 160/pixel), with roughly solar metallicities and spaced evenly in \teff and \Rstar. When presented with an unknown target spectrum, we find the best reference spectrum by performing the cross-correlation procedure described above on a single order of the target spectrum. The reference spectrum which gives the largest median cross-correlation peak with the target is deemed to have the greatest similarity to the target and is then used to shift the rest of the target spectrum. We selected the spectral order that spans $\lambda=5120-5210$~\angstrom as the benchmark order due to the rich spectral information contained in this region across all spectral types. In Figure~\ref{fig:bootstrap_spectrum}, we illustrate the successful registration of Barnard's star (GL699, spectral type M4V) to the NSO scale with our bootstrapping approach.

\subsection{Matching} 
\label{ssec:matching}
Once the target spectrum has been shifted and flattened onto the library wavelength scale, we compare it to each of the library spectra, in segments of 100~\angstrom (see table \ref{table:wl_regions}). In each comparison, we first modify the library spectrum by applying a rotational broadening kernel (Equation 17.12 in \citealt{Gray92}) to account for the relative \vsini between the target and reference stars. We found that it was necessary to set a maximum relative \vsini limit, chosen to be 10~\kms, to prevent the reference spectrum from being broadened excessively. If the kernel was allowed to go to arbitrarily high \vsini values, it would allow seemingly better matches even with dissimilar spectra due to the loss of spectral information from broadening. We also fit a cubic spline through the residuals between the target and reference spectra, using knots placed at 20~\angstrom intervals. Subtracting out this spline amounts to a high-pass filter which ensures that the \chisq is not influenced by slowly-varying ($\gtrsim  20$~\angstrom) residual structure due to imperfections in the blaze function removal.

We adopt the unnormalized $\chi^2$ statistic the figure of merit that quantifies degree of similarity between the target and modified reference spectra:
\begin{displaymath}
	\chi^2 = \sum(s_{ref} - s_{target})^2.
\end{displaymath}
The pixel-by-pixel Poisson uncertainties are not included because the main residual differences in the spectra are due to astrophysical differences, not photon noise. Including photon errors would give lower \chisq values when comparing with library spectra with poorer signal-to-noise ratios, even if these spectra are dissimilar to the target. We verify in Section~\ref{ssec:precision} that \SpecMatch is dominated by systematics, not photon statistics down to SNR of 10/pixel.

\begin{figure*}
    \epsscale{1.2}
	\plotone{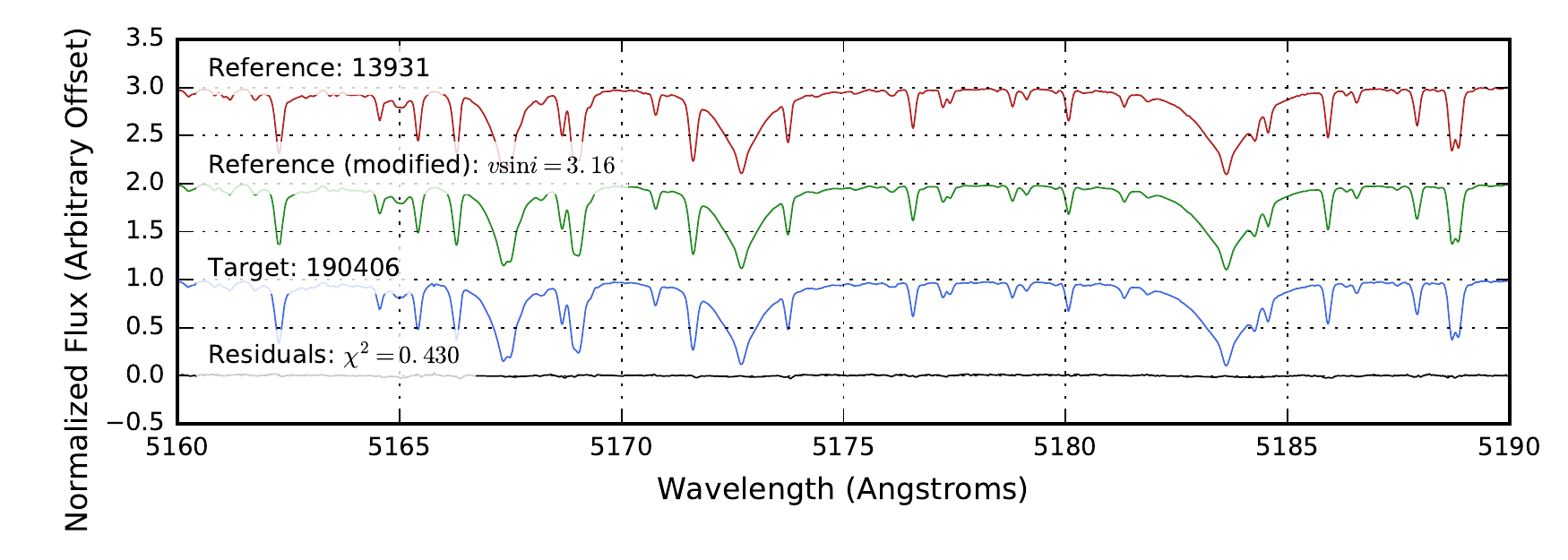}
	\epsscale{1.2}
	\plotone{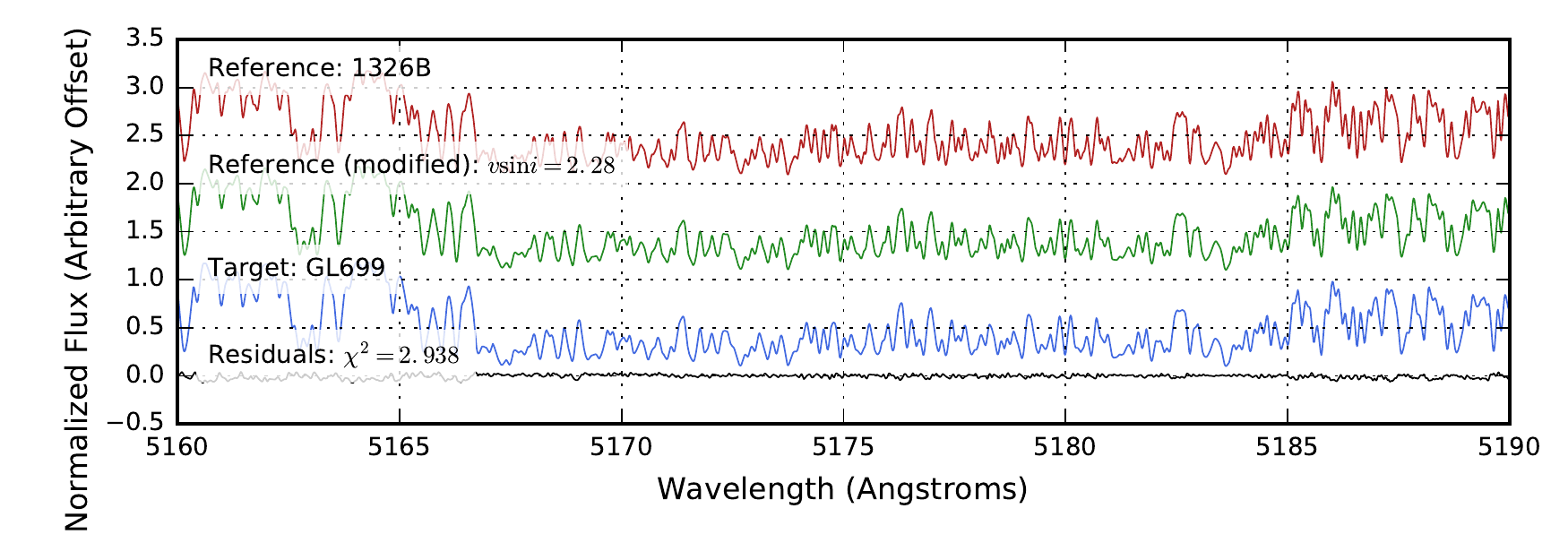}
	\caption{Best matching library spectra to HD190406 (top) and GL699 (bottom). The modified reference spectra are the library spectra after applying a broadening kernel and fitting a cubic spline to the continuum (see \autoref{ssec:matching}).\label{fig:best_match_spectra}} 
\end{figure*}

\begin{figure*}
	\epsscale{1.2}
	\plotone{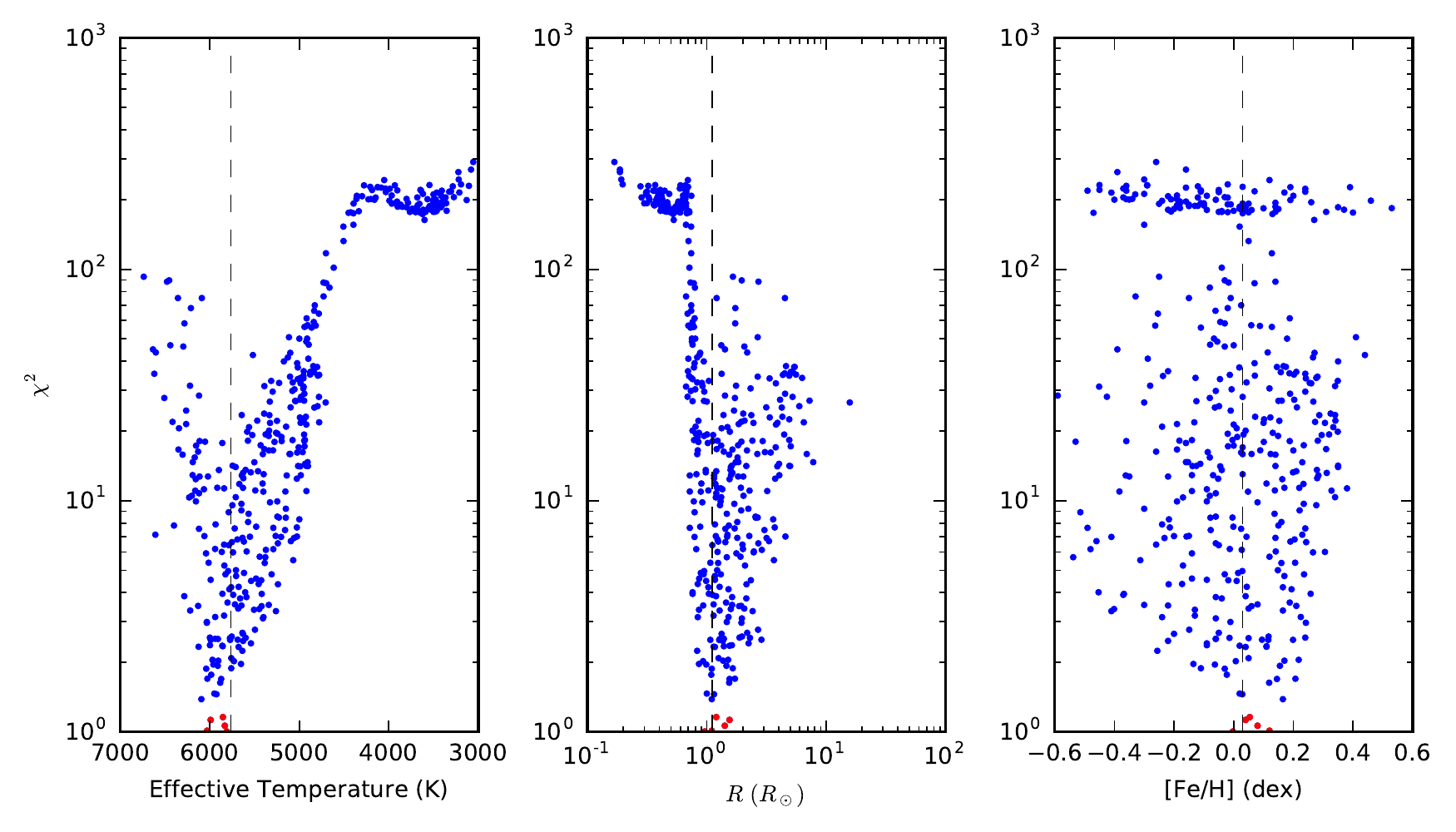}
	\epsscale{1.2}
	\plotone{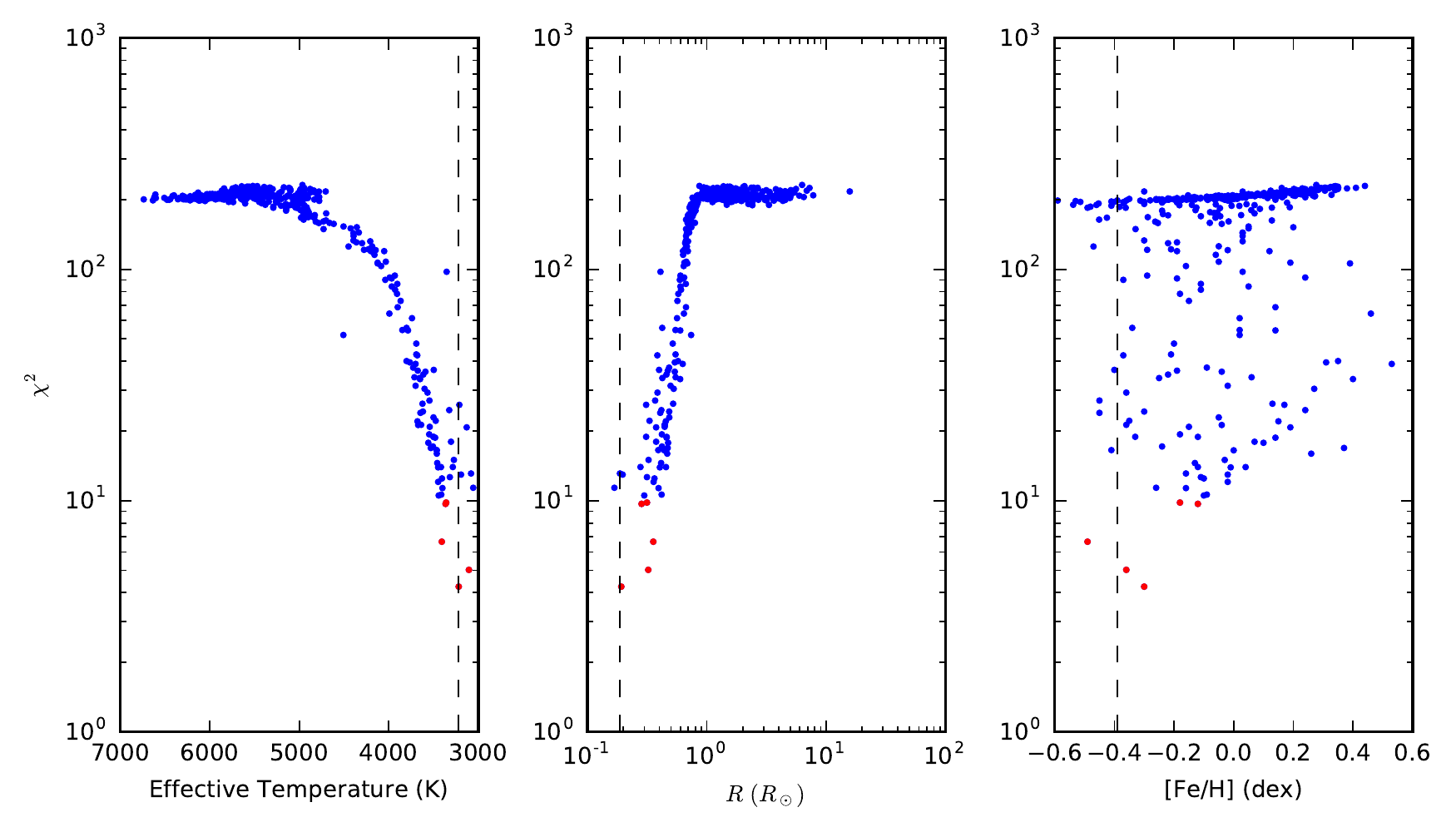}
	\caption{Best chi-squared values for the matches between the spectrum of HD190406 (top) and GL699 (bottom) with each library spectrum, plotted against the library star parameters (see \autoref{ssec:matching}). The stars with the five lowest chi-squared values are displayed in red, and we see that there is a sharp minimum in chi-squared close to the true target parameters, indicated with a vertical line. This match was performed in the wavelength region 5300 - 5400 \angstrom. \label{fig:chi_squared_surface}}
\end{figure*}

Allowing for \vsini and continuum normalization to float as free parameters, we use a non-linear least-squares minimization package \texttt{lmfit} \citep{lmfit} to minimize \chisq, finding the best possible match between the target and reference spectra. We then repeat this matching process over all the library spectra, recording in each case the lowest \chisq value achieved. We illustrate the results of the matching step for a HD190406, a solar-type star, and Barnard's star (GL699), an M4V star in Figures~\ref{fig:best_match_spectra} and \ref{fig:chi_squared_surface}. Figure~\ref{fig:best_match_spectra} shows the best-matching spectra from the library and Figure~\ref{fig:chi_squared_surface} shows the distribution of \chisq.

\begin{table}[h]
\centering
\begin{tabular}{cc}
\tableline
\tableline
Region & Wavelength Range  \\
\tableline
1 	& 5000 -- 5100 \angstrom\\
2 	& 5100 -- 5200 \angstrom\\
3 	& 5200 -- 5300 \angstrom\\
4 	& 5300 -- 5400 \angstrom\\
5 	& 5400 -- 5500 \angstrom\\
6	& 5500 -- 5600 \angstrom\\
7 	& 5600 -- 5700 \angstrom\\
8	& 5700 -- 5800 \angstrom\\
\tableline
\tableline
\end{tabular}
\caption{Wavelength regions used in both the matching and linear combination steps in \SpecMatch. \label{table:wl_regions}}
\end{table}

\subsection{Linear Combination} 
\label{ssec:lincomb}
The \SpecMatch routine then interpolates between the parameters of the library spectra by synthesizing linear combinations of the five best-matching spectra as found in the previous step. We form a new composite spectrum, $S_{lc} = \sum_{i=1}^{5} c_i \cdot S_i$, where each spectrum, $S_i$, is broadened by the optimal \vsini found in the previous matching stage. We chose to use five spectra in these linear combinations by trial and error. We use \texttt{lmfit} to find the set of coefficients $\{c_1, c_2, ..., c_5\}$ which minimizes \chisq when compared with the target spectrum. The $c_i$'s are subject to the constraint that they should sum to unity by incorporating a narrow Gaussian prior of width $\sigma = 10^{-3}$, such that \chisq:
\begin{displaymath}
	\chi^2 = \sum(s_{lc} - s_{target})^2 + \left(\sum c_i - 1\right)^2 / 2\sigma^2.
\end{displaymath}
In this step, we continue to correct for differences in continuum normalization by fitting a cubic spline and allowing the spline parameters to float as we minimize \chisq.

\begin{figure*}
	\plotone{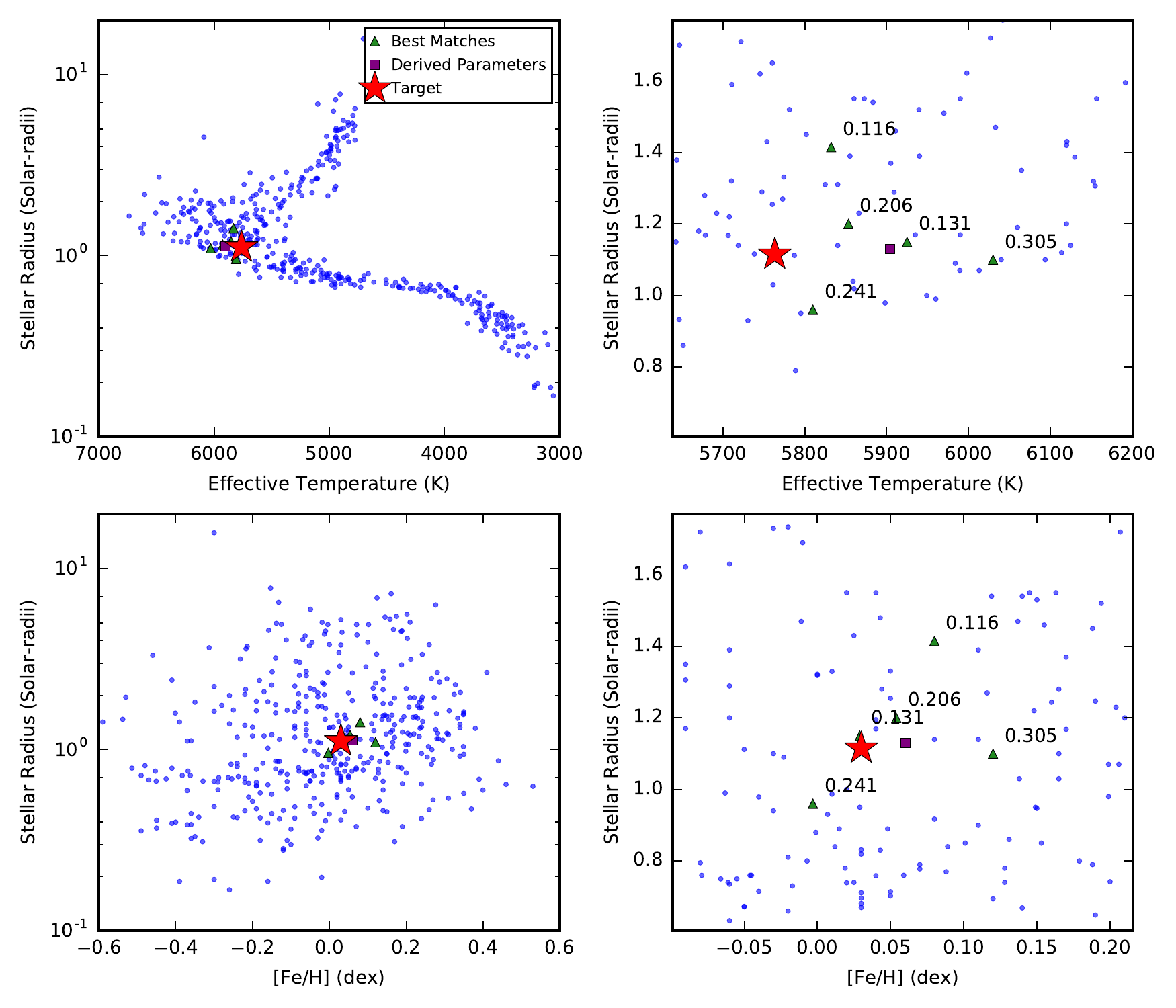}
	\plotone{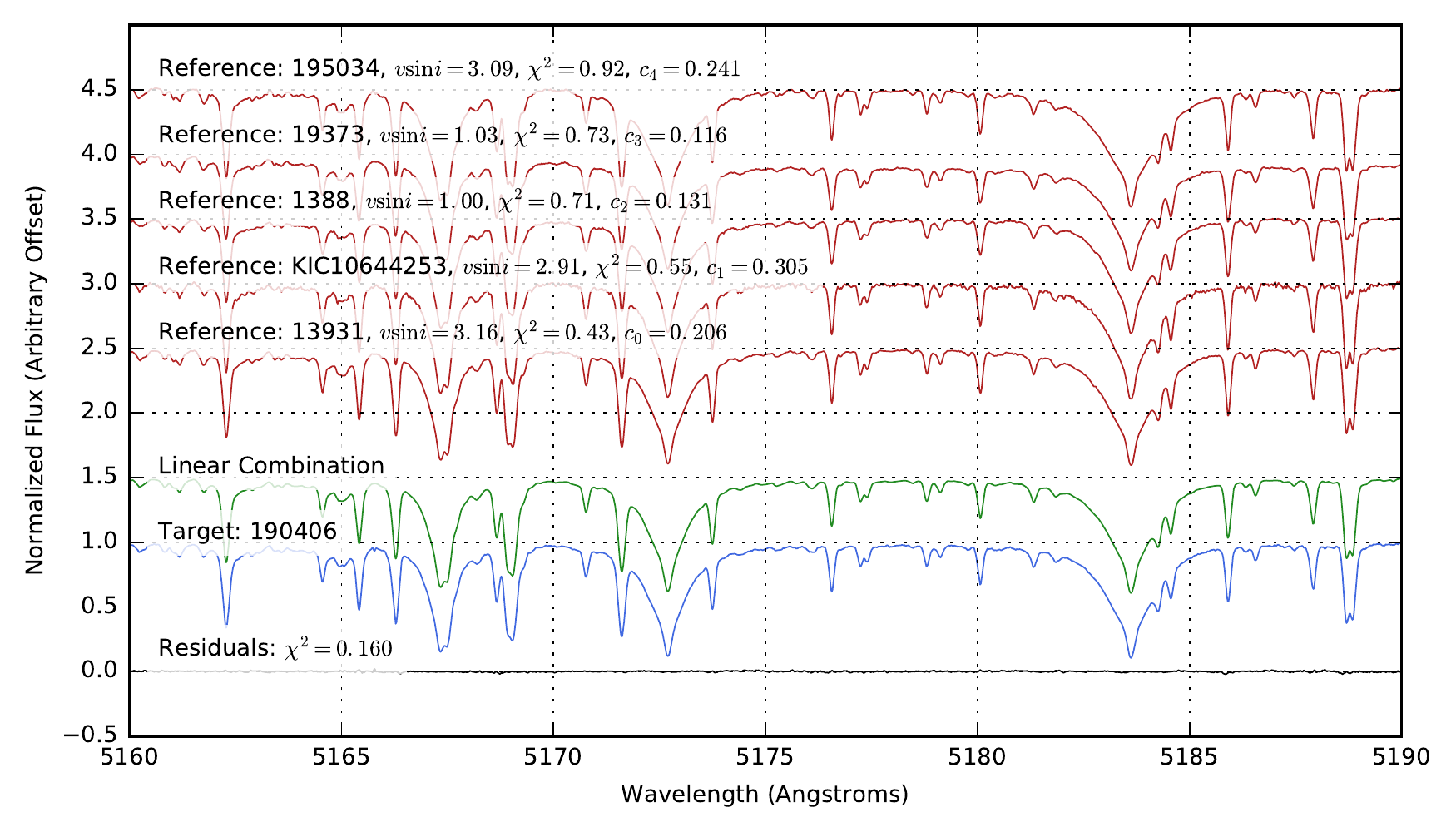}
	\caption{Results from the linear combination approach (\autoref{ssec:lincomb}) for the star HD190406. Top: The position of the five reference stars used, with their respective coefficients. The star indicates the library parameters of the target, while the purple square denotes the weighted average of the reference parameters. Bottom: The reference spectra and linear combination found. The final \chisq of 0.160 for the linear combination is lower than the \chisq for the best matching single spectrum, \chisq = 0.43.\label{fig:Gstar_lincomb}}
\end{figure*}

\begin{figure*}
	\plotone{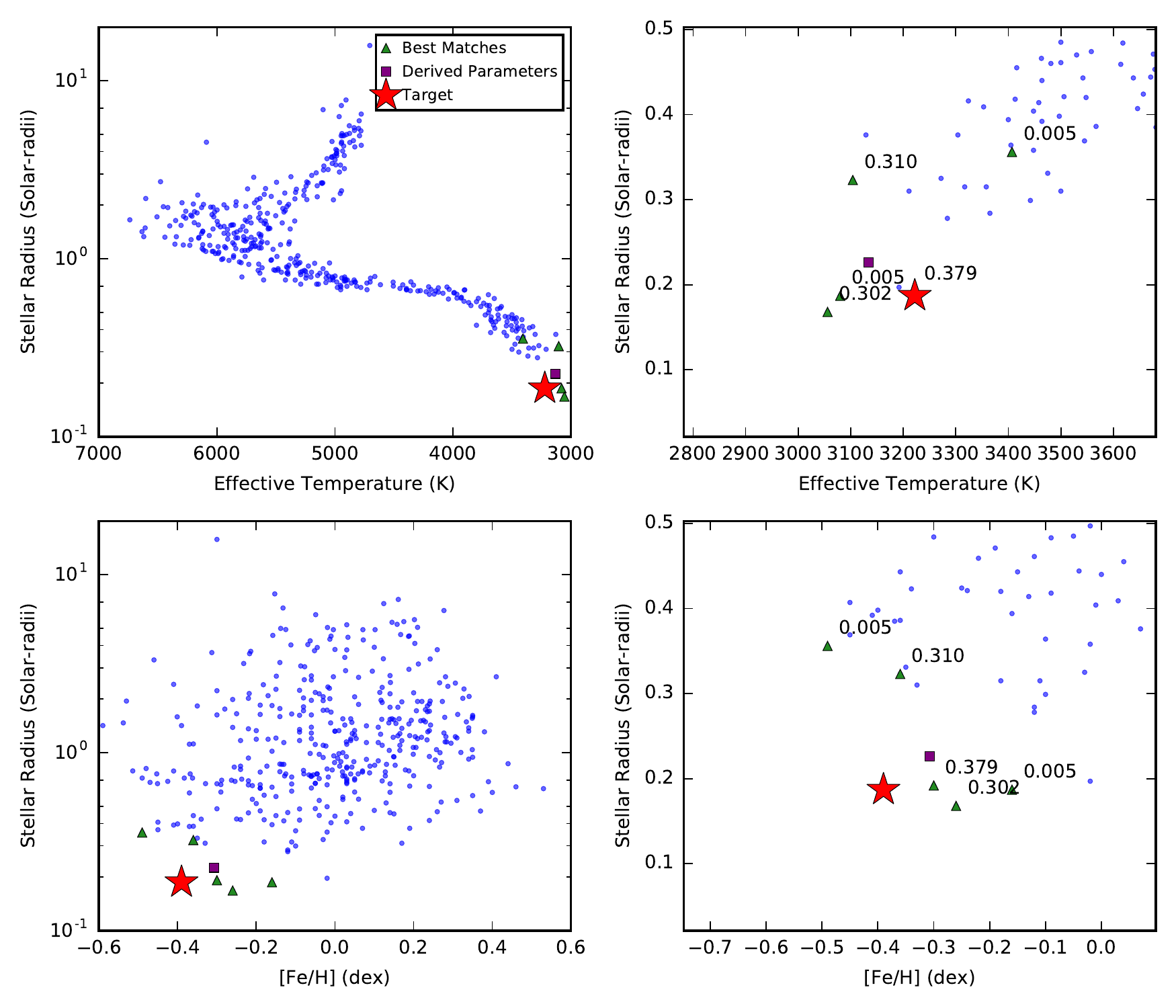}
	\plotone{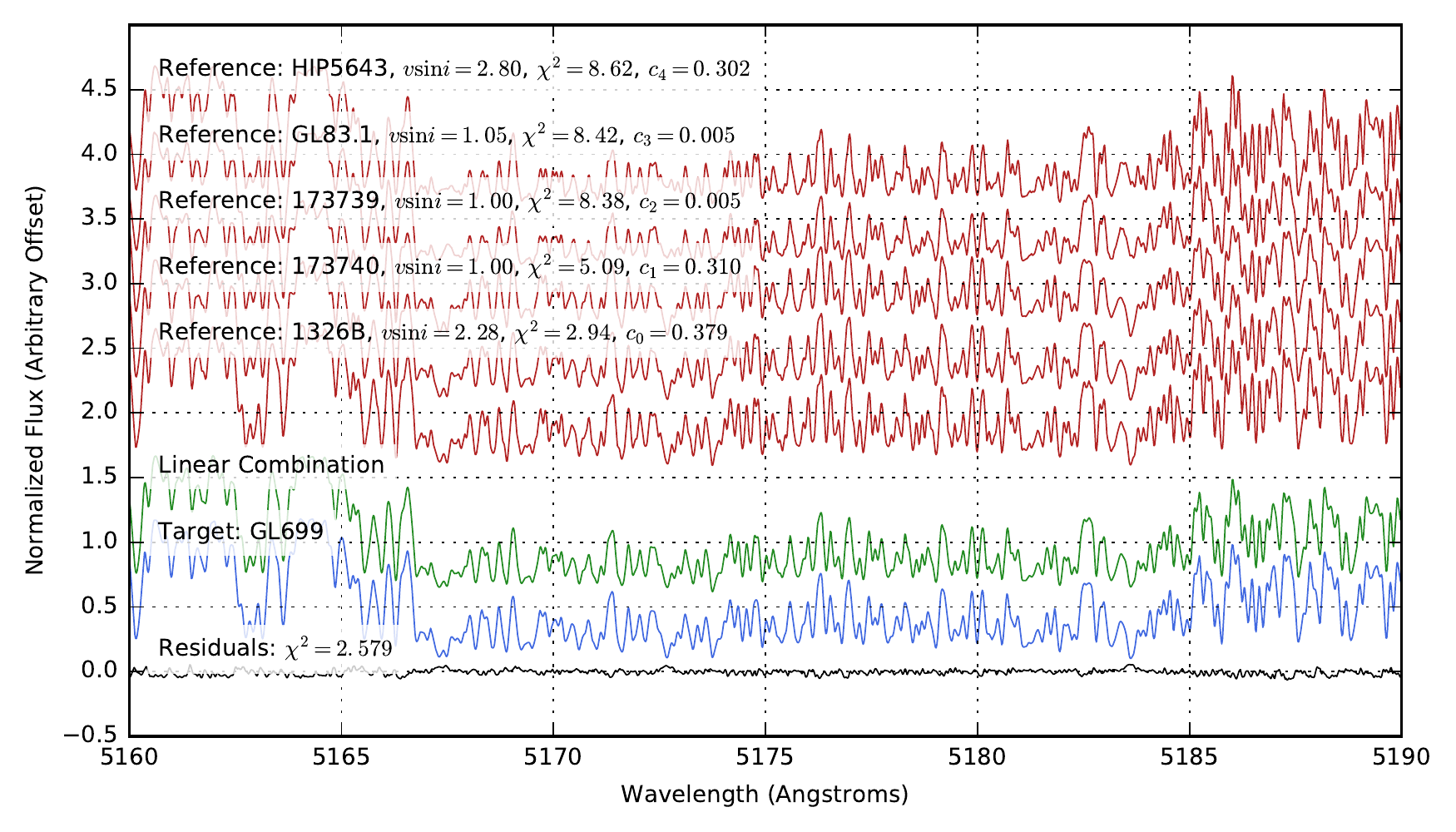}
	\caption{Same as Figure~\ref{fig:Gstar_lincomb}, but for the star GL699. The final residuals between the target star and linear combination vanish almost completely (see \autoref{ssec:lincomb}). \label{fig:Mstar_lincomb}}
\end{figure*}

We thus obtain a new spectrum which matches the target spectrum even more closely than any individual library spectrum. We show example results in Figures~\ref{fig:Gstar_lincomb} and \ref{fig:Mstar_lincomb}.

The set of coefficients $c_i$ found which minimizes \chisq is then used to create a weighted average of the parameters of the reference stars. In order to incorporate the spectral information from the entire spectrum, we average the derived parameters from each 100 \angstrom segment to obtain a final set of stellar parameters for the target. In total, it takes approximately 10 minutes to obtain the final parameters from a raw, unshifted spectrum on a modern (c. 2016) desktop computer.

\section{Performance}
\label{sec:performance}

\subsection{Accuracy}
\label{ssec:accuracy}
To assess the accuracy of the \SpecMatch, we performed an internal cross-validation, where we treated each star in the library as an unknown target and ran \SpecMatch to compute their parameters from the remaining library spectra. We then compared these derived parameters to their library values. The difference between the derived parameters and the library parameters reflects the errors in the \SpecMatch algorithm and in the library parameters.

\begin{figure*}[t]
\epsscale{1.2}
\plotone{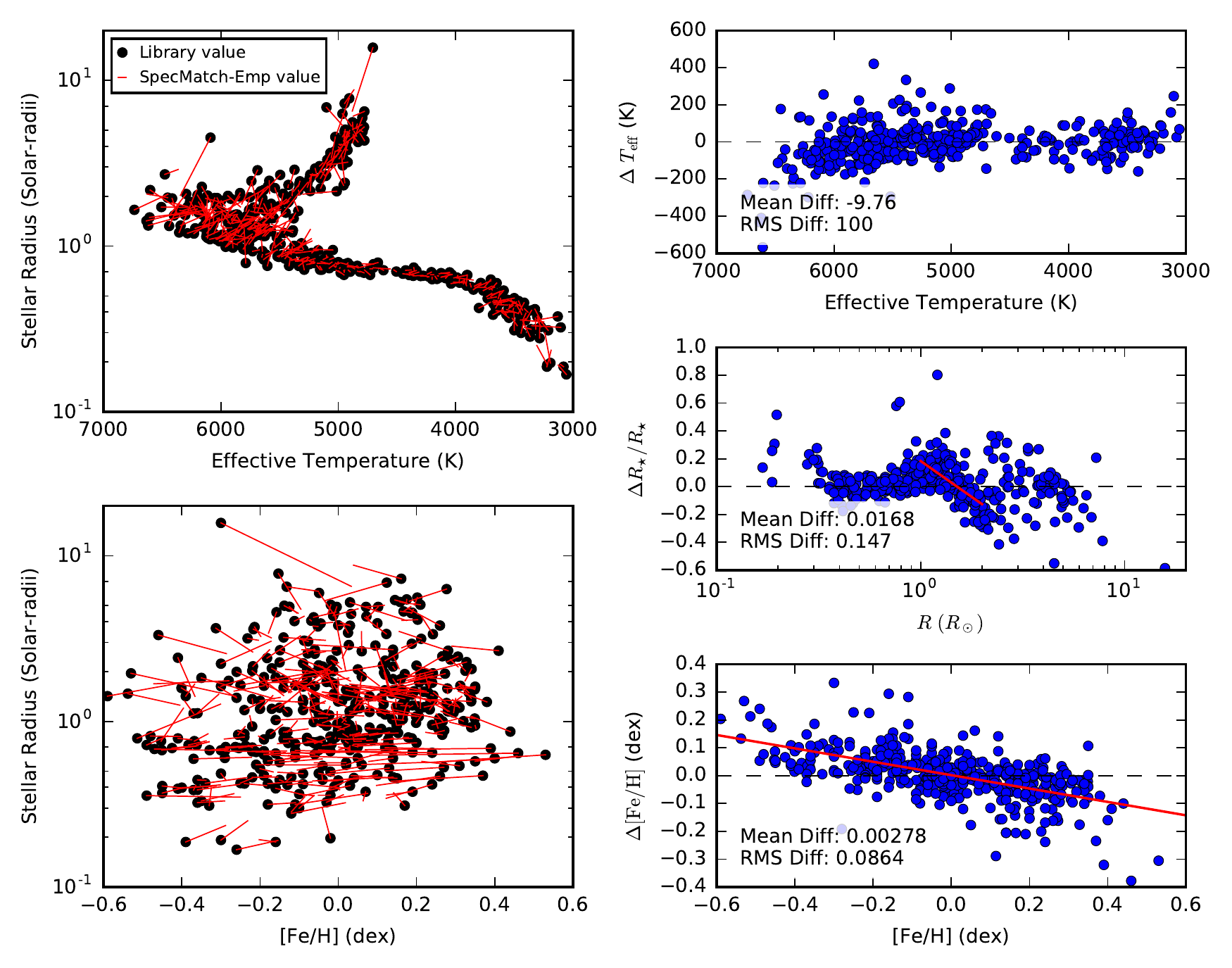}
\caption{Comparison of library parameters to SpecMatch-derived parameters for each library spectrum in the validation process (\autoref{ssec:accuracy}). Left: Black points indicate the library stellar parameters, while red lines point to the SpecMatch parameters. Right: Differences between the library values of \teff, \Rstar, \fe and the derived values. The red lines show the trends in the residuals which we attempt to fit out. \label{fig:five_pane_nodetrend}} 
\end{figure*}

Figure \ref{fig:five_pane_nodetrend} shows the results of this validation process. We notice a general tendency for the residuals to be most negative for larger values of the derived parameter, and most positive for smaller values. This can be partly explained by the fact that our library spectra are not on an infinite grid, but occupy only a finite region of parameter space. Target stars must necessarily match stars on the interior of that space, resulting in their derived parameters pulled toward the interior of the distribution. On the one hand, this effect amounts to a source of systematic uncertainty resulting from the use of real spectra of nearby stars, as opposed to using model spectra which can be synthesized with arbitrary properties. On the other hand, \SpecMatch guards against returning combinations of parameters that are not realized among nearby stars, whereas spectral synthesis may wander into unphysical parameter space.

We attempt to  mitigate the regression toward the mean effect by recalibrating the derived parameters. The effect is most pronounced for the residuals in \fe, and we fit a first order polynomial to the residuals, obtaining the following correction:
\begin{equation} 
\label{eq:fe_correction}
	\fe_{cal} = 1.240 \fe_{SM} - 0.0018
\end{equation}
We perform a similar detrending for \Rstar, but restrict the fit only to derived parameters of $1.0 < \Rstar / \Rsun < 2.0$, where the effect is most pronounced. For other values of \Rstar, the library stars are restricted to a narrow region of the HR diagram, so the derived parameters are also confined to a narrow range. However, the main sequence is relatively broad for for main sequence stars with $\teff > 5500$~K and \Rstar = 1--2~\Rsun, so we see a larger spread of derived \Rstar values and the regression to the mean effect is more pronounced.

The empirical relation we used to recalibrate the derived radii is given in Equation~\ref{eq:rad_correction}. In this case, the calibration relation is quadratic, as we are fitting out a trend in $\Delta R/R$.
\begin{equation} \label{eq:rad_correction}
	R_{\star, cal} = 0.560 R_{\star, SM}^2 + 0.165 R_{\star, SM}
\end{equation}

\begin{figure*}[t]
\epsscale{1.2}
\plotone{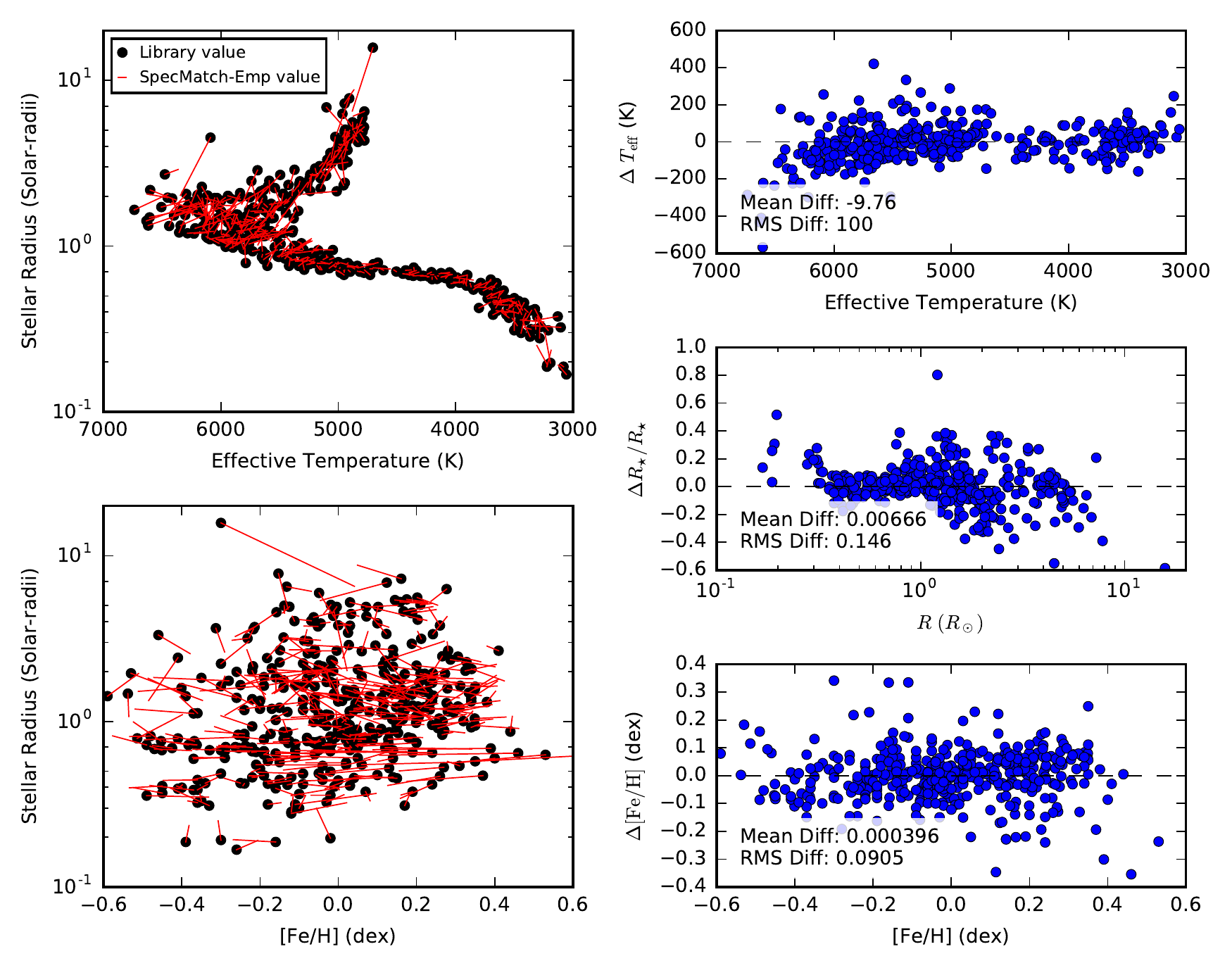}
\caption{Same as \ref{fig:five_pane_nodetrend}, but after performing detrending (see \autoref{ssec:accuracy}).\label{fig:five_pane}} 
\end{figure*}

The final, detrended results are shown in Figure~\ref{fig:five_pane}. For all our library stars, the differences between the derived and library values had a scatter of \sigteff in \teff, \sigRstar in \Rstar, and \sigfe in \fe. These values, listed in Table~\ref{table:specmatch_errs}, are the uncertainties which should be adopted for the output of \SpecMatch.

\begin{figure*}[t]
\epsscale{1.2}
\plotone{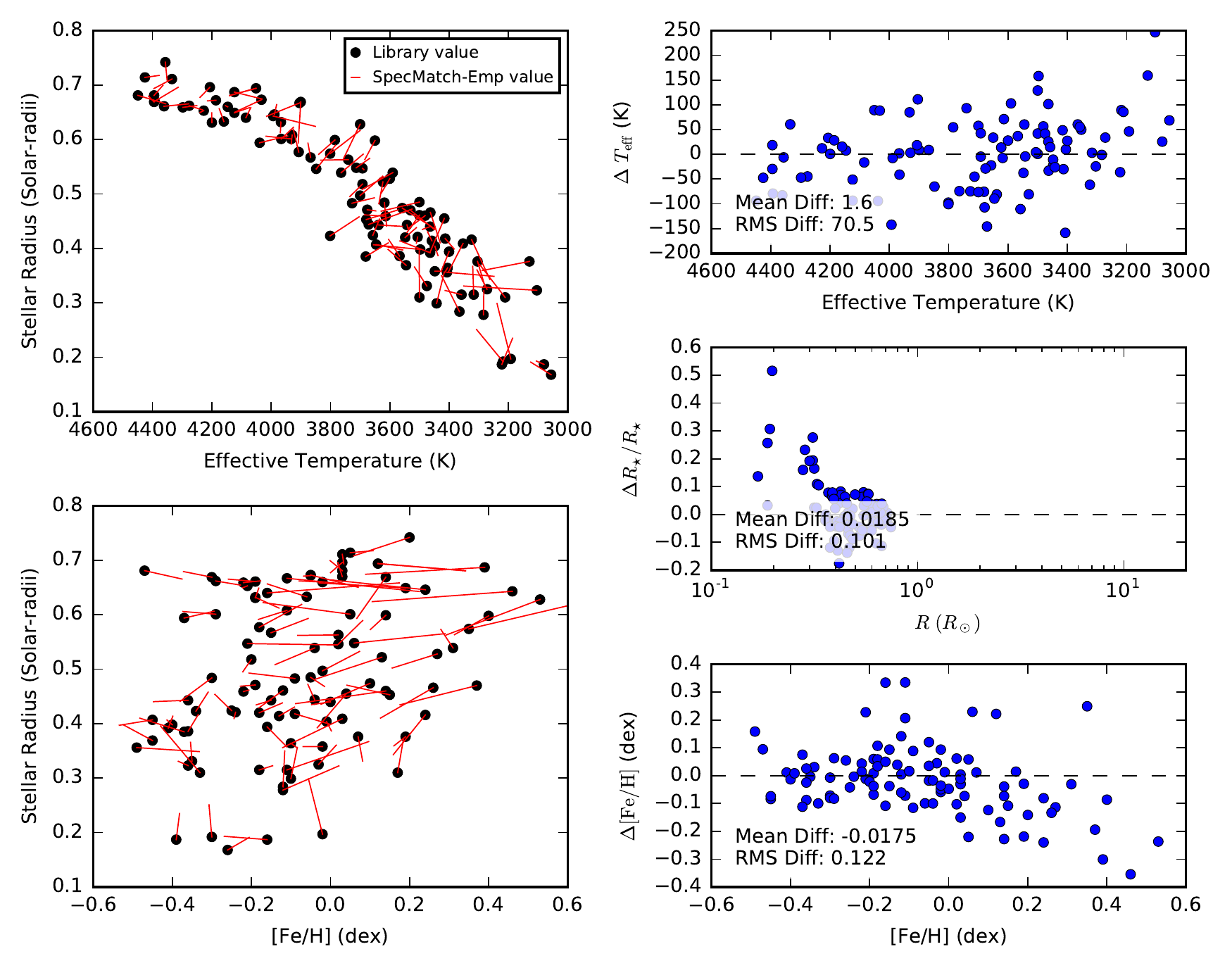}
\caption{Same as \ref{fig:five_pane}, but for the library stars with \teff < 4500 K. Performance is improved for this selection of stars due to the more limited spread in main sequence stellar parameters in the HR diagram (see \autoref{ssec:accuracy}). \label{fig:five_pane_cool}} 
\end{figure*}

When restricting our analysis to the cool stars (Figure \ref{fig:five_pane_cool}), with $\teff < 4500$~K, the performance of \SpecMatch improves to \sigteffcool in \teff, \sigRstarcool in \Rstar, and \sigfecool in \fe. For cool stars, most of the stars are from the \cite{Mann15} sample, which has a median uncertainty of \sigteffcool in \teff, \sigRstar in \Rstar and \sigfe in \fe. In this region then, the accuracy of SpecMatch-Emp of the routine appears to be primarily limited by the uncertainties in the library parameters, which set the theoretical accuracy limit.

\begin{table}[ht]
\centering
\begin{tabular}{cccc}
\tableline
\tableline
    & $\sigma(\teff)$ & $\sigma(\Delta\Rstar/\Rstar)$ & $\sigma(\fe)$ \\
	& K               &	\%	                          &  dex \\
\tableline
All stars 			& 100	& 15	& 0.09 \\
$\teff < 4500$~K	& 70	& 10	& 0.12 \\
$\teff \geq 4500$~K & 110	& 16	& 0.08 \\
\tableline
\tableline
\end{tabular}
\caption{RMS difference between \SpecMatch-derived and library parameters for each star. Different uncertainties may be adopted for stars from different regions of the HR diagram, as the accuracy of \SpecMatch depends on the distribution of library stars in that region as well as the scatter of the library parameters for those stars. \label{table:specmatch_errs}}
\end{table}

\subsection{Performance at Low SNR}
\label{ssec:precision}

We also investigated the effect of photon shot noise on the precision of the \SpecMatch derived parameters. We chose a subset of 22 stars from the library, distributed across the HR diagram and with original SNR > 160/pixel. Their unshifted spectra are then degraded to lower SNR by injecting Gaussian noise into the raw spectra. We chose target SNRs of 120, 80, 40, 20, an 10 per HIRES pixel. For each target spectrum and SNR level we generated 20 noisy spectra. We then analyzed these spectra through the \SpecMatch routine and compared the final properties to those derived from the original high-SNR spectra.

\begin{figure*}[t]
\epsscale{1.2}
\plotone{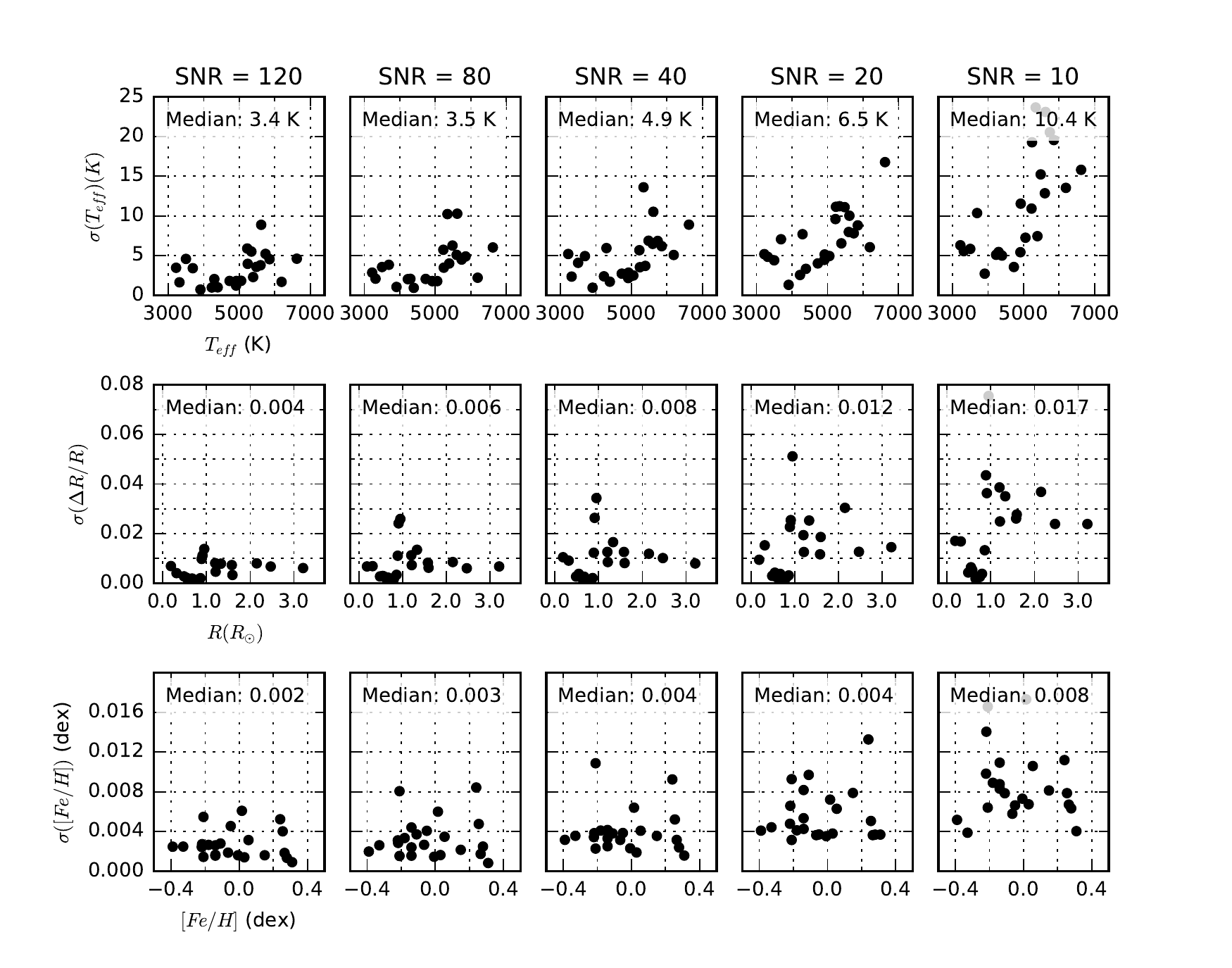}
\caption{Scatter of SpecMatch-derived parameters as a function of SNR of the target spectrum, from the noise study described in \autoref{ssec:precision}. Each point represents the median RMS difference between the parameters derived from 20 noisy spectra, and the derived parameter of the original, high SNR spectrum. As SNR decreases, the median scatter increases, and is representative of the effect of photon noise on the precision of \SpecMatch. The decrease in precision is most pronounced for stars with greater \teff and radius, as these stars have fewer spectral lines and are thus most susceptible to features being washed out by noise.\label{fig:noise_study}} 
\end{figure*}

Figure~\ref{fig:noise_study} shows the RMS difference between the noisy derived parameters and the high-SNR derived parameter, for each star and target SNR. Treating the high-SNR parameter as the ground truth, these results are representative of the random errors of the derived parameters caused by noise in the input spectrum. As expected, the median scatter increases as the SNR decreases. Nonetheless, the median scatter is only 10.4 K in \teff, 1.7 \% in \Rstar, and 0.008 dex in \fe even at an SNR~=~10/pixel. These are significantly smaller than the algorithmic limitations in accuracy from the matching process, demonstrating the robustness of the \SpecMatch algorithm even at low SNRs. Table~\ref{table:noise_scatter} lists the scatter in each parameter at different SNRs.

Furthermore, we note that the increase in scatter with noise is greater for stars with higher \teff and radius. These stars have fewer spectral lines, so random noise has a bigger impact on the derived parameters. The cooler, small stars have much more spectral information in the wavelength region, and so \SpecMatch is more robust to noise for these spectra.

\begin{table}[ht]
\centering
\begin{tabular}{cccc}
\tableline
\tableline
    & $\sigma(\teff)$ & $\sigma(\Delta\Rstar/\Rstar)$ & $\sigma(\fe)$ \\
SNR & K               &		                          &  dex \\
\tableline
120 & 3.4	& 0.004	& 0.002 \\
80	& 3.5	& 0.006	& 0.003 \\
40 	& 4.9	& 0.008	& 0.004 \\
20	& 6.5	& 0.012	& 0.004 \\
10	& 10.4	& 0.017	& 0.008 \\
\tableline
\tableline
\end{tabular}
\caption{RMS scatter in derived parameters at different signal-to-noise ratios. \label{table:noise_scatter}}
\end{table}

\subsection{Performance at Low Spectral Resolution}
\label{ssec:resolution}

The library spectra were observed with uniform spectral resolution of $R\approx 60000$. Here, we investigate the effect of lower spectral resolutions on the accuracy of the derived parameters. As spectral resolution decreases and narrow lines become washed out, we expect the performance of \SpecMatch to worsen since the algorithm relies on matching large numbers of spectral lines against the library spectra.

To perform this study, we use the same subset of 22 stars as in Section \ref{ssec:precision} and simulate lower spectral resolution by convolution with a Gaussian kernel. The degraded spectra were then passed through the \SpecMatch routine and a final set of properties was obtained.

\begin{figure*}[t]
\epsscale{1.2}
\plotone{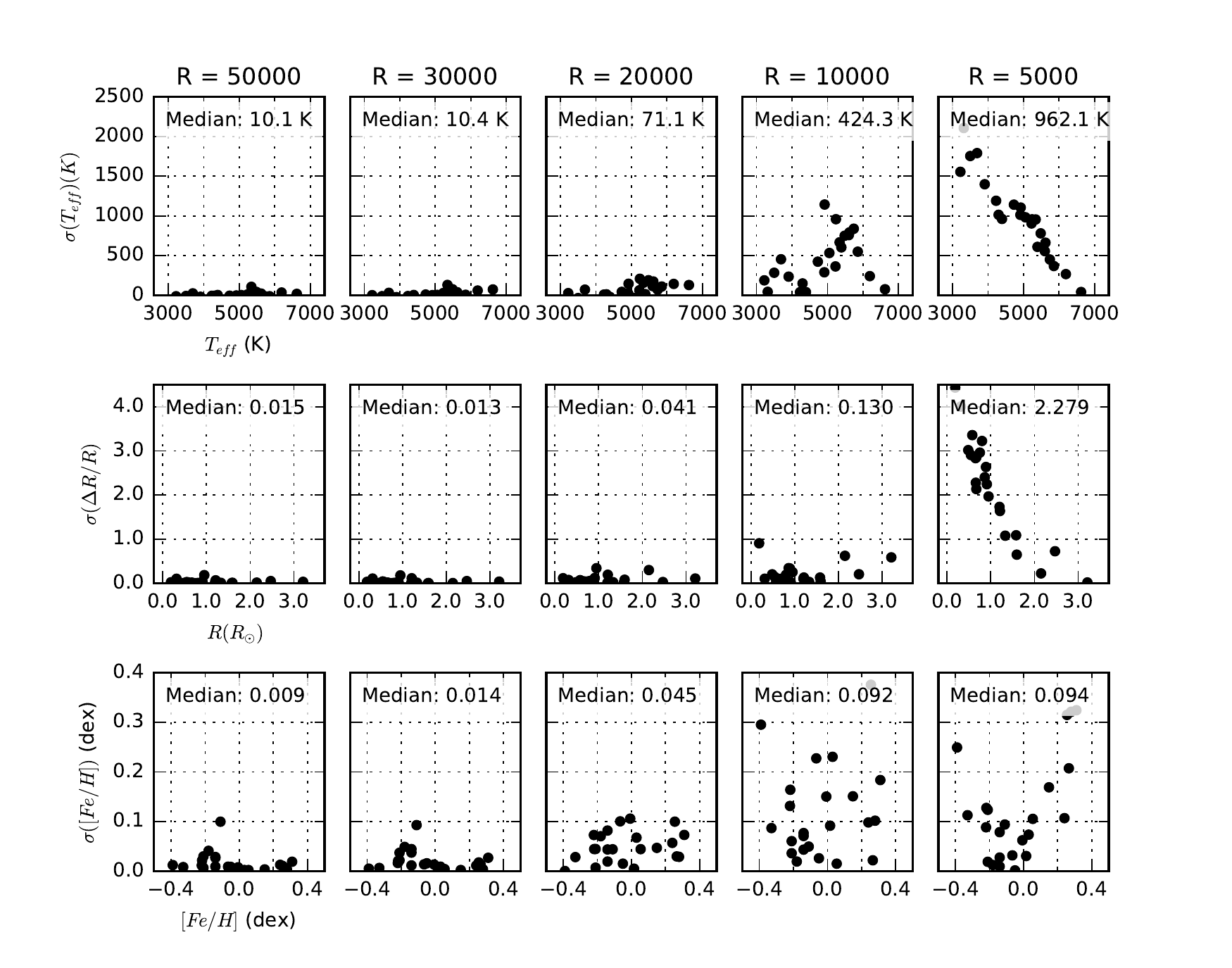}
\caption{Absolute difference between the parameters derived from the spectra with degraded spectral resolution and those derived from the original high-resolution spectra, analogous to figure \ref{fig:noise_study}. \SpecMatch remains accurate down to $R\approx30000$, beyond which the performance degrades rapidly (see \autoref{ssec:resolution}).\label{fig:res_study}} 
\end{figure*}

Just as in the previous section, we treat the properties obtained using the original high-resolution ($R\approx60000$) spectra as the ground truth. We plot the absolute difference between these properties and the derived properties from each degraded spectrum in Figure \ref{fig:res_study}. The scatter remains small down to $R = 30000$, with a median of 10 K in \teff, 1.3\% in \Rstar, and 0.014 dex in \fe. At $R = 20000$, the scatter due to the decreased resolution becomes comparable in size to the algorithmic uncertainties determined through our cross validation study (Section \ref{ssec:accuracy}). At even lower resolutions, \SpecMatch is no longer usable, producing very large errors particularly in \teff and \Rstar. Table \ref{table:res_scatter} summarizes these results.

\begin{table}[ht]
\centering
\begin{tabular}{cccc}
\tableline
\tableline
    & $\sigma(\teff)$ & $\sigma(\Delta\Rstar/\Rstar)$ & $\sigma(\fe)$ \\
R & K               &		                          &  dex \\
\tableline
50000   & 10.1	& 0.015	& 0.009 \\
30000	& 10.4	& 0.013	& 0.014 \\
20000 	& 71.1	& 0.041	& 0.045 \\
10000	& 424	& 0.13	& 0.092 \\
5000	& 962	& 2.28	& 0.094 \\
\tableline
\tableline
\end{tabular}
\caption{Median scatter in derived parameters at different spectral resolutions. \label{table:res_scatter}}
\end{table}

A closer analysis of the intermediate results reveals that the spectral registration step is fairly insensitive to lower spectral resolution. The same shift results, accurate to approximately one pixel, are obtained for most orders even at $R = 10000$.

The performance of \SpecMatch suffers during the matching step. In the current implementation of \SpecMatch, we do not account for the spectrometer instrumental profile. As a resut, at lower resolution, the algorithm favors increased rotational broadening to account for the broader lines. This compensation by higher rotational broadening works well down to $R\approx20000$, beyond which it fails for two reasons. First, the kernels associated with rotational broadening and the instrumental profile are different. Second, during the matching step, we do not allow the broadening kernel to exceed a \vsini > 10~\kms. At a resolution of $\sim10000$ the instrumental profile has a width of $\sim 13$~\kms, which is above the maximum \vsini allowed during our fitting. 

As a result, at spectral resolutions below 20000, no reference spectrum can be broadened sufficiently to simulate the wider point-spread function of the spectrometer, and the matching algorithm produces a large number of similarly poor matches. The \chisq surfaces no longer have sharp minima and the final parameters obtained are less accurate. This decrease in performance particularly affects stars with low \teff and \Rstar. These M and K dwarfs have many narrow but overlapping lines in their spectra (e.g. Figure \ref{fig:shifted_spec}), which get smeared out into a continuum as the resolution decreases, preventing accurate matching with the library spectra. The spectra from hotter stars have fewer but broader lines which remain visible even in lower resolution spectra.

While we have made no explicit treatment of of spectrometer instrumental broadening, the current implementation of \SpecMatch is relatively insensitive to spectral resolutions down to $R\approx30000$. Difficulties in modeling spectra with $R\lesssim20000$ could be addressed by proper treatment of the instrumental broadening profile. We welcome commmunity contributions in this respect.

\section{Conclusions} 
\label{sec:conclusion}
We have compiled a library of high resolution, high signal-to-noise optical spectra of \libnum touchstone stars with well-measured properties. The \SpecMatch routine is able to rapidly extract fundamental properties of a star from its optical spectrum by comparison against this library. The density of the library and the quality of the associated stellar parameters enables \SpecMatch to acheive accuracies of \sigteff in \teff, \sigRstar in \Rstar and \sigfe in \fe. By incorporating information from across a large wavelength range, the algorithm is relatively robust to photon noise and can readily applied to stellar spectra with signal-to-noise ratios as low as 10 per pixel. The algorithm also remains robust at lower spectral resolutions until about $R\sim20000$.

A key advantage of \SpecMatch is its accuracy for cool stars with $\teff < 4500$~K. By using empirical, rather than synthetic, spectra to create the library, we have bypassed the difficulties that existing spectral synthesis codes have in modeling the complex spectra of cool stars. Instead, the rich spectral information contained in these stars' spectra is an advantage during the matching process, as the greater number of features improves both the accuracy and precision of the derived parameters. We expect that \SpecMatch will be a valuable tool for efficiently characterizing large numbers of late-type stars. Such characterization work will be a key observational follow-up effort for future transit surveys such as {\em TESS} \citep{Ricker14} that have an emphasis on M-stars.

The applicability of \SpecMatch is only valid within the region of stellar parameter space spanned by our library stars, i.e. \SpecMatch is not applicable to rapidly-rotating stars, young stars, chemically peculiar stars, extremely metal-poor stars, etc. Furthermore, the routine's accuracy is limited by the quality of the input parameters. Because the library stars are drawn from the sample of nearby stars, we anticipate that their parameters will be refined with future observations. The \SpecMatch library may be easily updated to include these updated parameters, with no modifications to the algorithm.

\acknowledgments
We thank Tabetha Boyajian, John Brewer, Debra Fischer, Eric Gaidos, Andrew Howard, Howard Isaacson, Heather Knutson, Andrew Mann, and Geoffrey Marcy for enlightening conversations that improved this manuscript. We thank the many observers who collected the Keck/HIRES data used here including R. Paul Butler, Debra A. Fischer, Benjamin J. Fulton, Andrew W. Howard, Howard Isaacson, John A. Johnson, Geoffrey W. Marcy, Kathryn M. G. Peek, Benjamin J. Fulton, Steven S. Vogt, Lauren M. Weiss, and Jason T. Wright. S.~W.~Y.\ acknowledges support through the Caltech Summer Undergraduate Research Fellowship program.  E.~A.~P.\ acknowledges support from a Hubble Fellowship grant HST-HF2-51365.001-A awarded by the Space Telescope Science Institute, which is operated by the Association of Universities for Research in Astronomy, Inc.~for NASA under contract NAS 5-26555. This work has made use of data from the European Space Agency (ESA) mission \Gaia (\url{http://www.cosmos.esa.int/gaia}), processed by the \Gaia Data Processing and Analysis Consortium (DPAC, \url{http://www.cosmos.esa.int/web/gaia/dpac/consortium}). Funding for the DPAC has been provided by national institutions, in particular the institutions participating in the \Gaia Multilateral Agreement. Some of the data presented herein were obtained at the W.~M.~Keck Observatory, which is operated as a scientific partnership among the California Institute of Technology, the University of California and the National Aeronautics and Space Administration. The Observatory was made possible by the generous financial support of the W.~M.~Keck Foundation. Finally, the authors wish to recognize and acknowledge the very significant cultural role and reverence that the summit of Maunakea has always had within the indigenous Hawaiian community.  We are most fortunate to have the opportunity to conduct observations from this mountain.

\clearpage
\appendix
\section{Library Stars}

\begin{deluxetable}{lrrrrrrrrr}
\tablecaption{SpecMatch Library \label{table:library_stars}}
\tabletypesize{\scriptsize}
\tablecolumns{5}
\tablewidth{0pt}
\tablehead{
	\colhead{Name} & 
	\colhead{\teff} &
	\colhead{\Rstar} & 
	\colhead{\logg} &
	\colhead{\fe} & 
    \colhead{\Mstar} & 
    \colhead{\logage} &
    \colhead{\plx} &
   	\colhead{$V$} &
    \colhead{Notes} \\
    \colhead{} & 
	\colhead{K} &
	\colhead{\Rsun} & 
	\colhead{dex} &
	\colhead{dex} & 
    \colhead{\Msun} & 
    \colhead{} &
    \colhead{mas} &
    \colhead{mag} &
    \colhead{}
}
\startdata
HD 100623     & 5140 $\pm$ 60     & 0.720 $\pm$ 0.040     & 4.56 $\pm$ 0.05     & -0.36 $\pm$ 0.05     & 0.75 $\pm$ 0.04     & 9.81 $\pm$ 0.50     & 104.6     & 6.0     & A1 \\
HD 101904     & 5883 $\pm$ 60     & 1.540 $\pm$ 0.190     & 4.10 $\pm$ 0.05     & 0.12 $\pm$ 0.05     & 1.12 $\pm$ 0.09     & 9.80 $\pm$ 0.11     & 12.5     & 8.2     & A1 \\
HD 102195     & 5277 $\pm$ 60     & 0.850 $\pm$ 0.050     & 4.50 $\pm$ 0.05     & 0.10 $\pm$ 0.05     & 0.89 $\pm$ 0.04     & 9.74 $\pm$ 0.49     & 34.0     & 8.1     & A1 \\
HD 102444     & 5227 $\pm$ 60     & 2.670 $\pm$ 0.390     & 3.60 $\pm$ 0.05     & -0.00 $\pm$ 0.05     & 1.02 $\pm$ 0.21     & 9.98 $\pm$ 0.27     & 8.5     & 8.0     & A1 \\
HD 102956     & 4985 $\pm$ 60     & 3.790 $\pm$ 0.470     & 3.38 $\pm$ 0.05     & 0.26 $\pm$ 0.05     & 1.30 $\pm$ 0.15     & 9.70 $\pm$ 0.17     & 8.1     & 7.9     & A1 \\
HD 103459     & 5722 $\pm$ 60     & 1.710 $\pm$ 0.210     & 4.03 $\pm$ 0.05     & 0.24 $\pm$ 0.05     & 1.15 $\pm$ 0.12     & 9.81 $\pm$ 0.15     & 16.0     & 7.6     & A2 \\
HD 103890     & 6120 $\pm$ 60     & 1.430 $\pm$ 0.160     & 4.19 $\pm$ 0.05     & 0.03 $\pm$ 0.05     & 1.17 $\pm$ 0.07     & 9.63 $\pm$ 0.09     & 13.7     & 8.0     & A1 \\
HD 104437     & 5761 $\pm$ 60     & 1.030 $\pm$ 0.090     & 4.40 $\pm$ 0.05     & 0.17 $\pm$ 0.05     & 1.03 $\pm$ 0.04     & 9.64 $\pm$ 0.35     & 16.6     & 8.6     & A1 \\
\enddata
\tablecomments{Properties of stars in the \SpecMatch library assembled from literature sources and this work.}
\tablenotetext{A}{\cite{Brewer16} sample (\autoref{sec:sample_brewer}). \teff, \logg, \fe from single line fits to stellar models; \Mstar, \Rstar, age from \isochrones.}
\tablenotetext{B}{Asteroseismic sample from \cite{Bruntt12} (\autoref{sec:sample_asteroseismic}). \logg from asteroseismic analysis; \teff and \fe from individual spectral line fits; \Mstar, \Rstar and age from \isochrones.}
\tablenotetext{C}{New K-dwarf sample (\autoref{ssec:kdwarfs}). \teff from empirical V-K color-temperature relationship; \Rstar from combining \teff with parallax and \Fbol calculated from SED fitting; \fe from SME fitting in \cite{Gaidos13}; \Mstar, \logg, age from \isochrones.}
\tablenotetext{D}{\cite{Mann15} sample (\autoref{sec:sample_mann}). \teff from stellar atmosphere models; \Rstar from combining \teff with \Fbol obtained from flux-calibrated spectra; \fe and \Mstar from empirical relationships; \logg calcualted directly from \Rstar and \Mstar; age from \isochrones.}
\tablenotetext{E}{Interferometric sample, as compiled in \cite{VonBraun14} (\autoref{sec:sample_interferometric}). \Rstar from optical and infrared interferometry; \teff and \fe by a variety of methods; \Mstar, \logg and age from \isochrones.}
\tablenotetext{1}{Parallaxes from \Gaia DR1 \citep{Gaia16a}.}
\tablenotetext{2}{Parallaxes from \Hipparcos \citep{VanLeeuwen07}.}
\tablenotetext{3}{No parallaxes found.}
\end{deluxetable}

\bibliography{manuscript.bib}

\begin{thebibliography}{}
\expandafter\ifx\csname natexlab\endcsname\relax\def\natexlab#1{#1}\fi

\bibitem[{{Auvergne} {et~al.}(2009){Auvergne}, {Bodin}, {Boisnard}, {Buey},
  {Chaintreuil}, {Epstein}, {Jouret}, {Lam-Trong}, {Levacher}, {Magnan},
  {Perez}, {Plasson}, {Plesseria}, {Peter}, {Steller}, {Tiph{\`e}ne}, {Baglin},
  {Agogu{\'e}}, {Appourchaux}, {Barbet}, {Beaufort}, {Bellenger}, {Berlin},
  {Bernardi}, {Blouin}, {Boumier}, {Bonneau}, {Briet}, {Butler}, {Cautain},
  {Chiavassa}, {Costes}, {Cuvilho}, {Cunha-Parro}, {de Oliveira Fialho},
  {Decaudin}, {Defise}, {Djalal}, {Docclo}, {Drummond}, {Dupuis}, {Exil},
  {Faur{\'e}}, {Gaboriaud}, {Gamet}, {Gavalda}, {Grolleau}, {Gueguen},
  {Guivarc'h}, {Guterman}, {Hasiba}, {Huntzinger}, {Hustaix}, {Imbert},
  {Jeanville}, {Johlander}, {Jorda}, {Journoud}, {Karioty}, {Kerjean},
  {Lafond}, {Lapeyrere}, {Landiech}, {Larqu{\'e}}, {Laudet}, {Le Merrer},
  {Leporati}, {Leruyet}, {Levieuge}, {Llebaria}, {Martin}, {Mazy}, {Mesnager},
  {Michel}, {Moalic}, {Monjoin}, {Naudet}, {Neukirchner}, {Nguyen-Kim},
  {Ollivier}, {Orcesi}, {Ottacher}, {Oulali}, {Parisot}, {Perruchot},
  {Piacentino}, {Pinheiro da Silva}, {Platzer}, {Pontet}, {Pradines},
  {Quentin}, {Rohbeck}, {Rolland}, {Rollenhagen}, {Romagnan}, {Russ}, {Samadi},
  {Schmidt}, {Schwartz}, {Sebbag}, {Smit}, {Sunter}, {Tello}, {Toulouse},
  {Ulmer}, {Vandermarcq}, {Vergnault}, {Wallner}, {Waultier}, \&
  {Zanatta}}]{Auvergne09}
{Auvergne}, M., {Bodin}, P., {Boisnard}, L., {et~al.} 2009, \aap, 506, 411

\bibitem[{{Baines} {et~al.}(2009){Baines}, {McAlister}, {ten Brummelaar},
  {Sturmann}, {Sturmann}, {Turner}, \& {Ridgway}}]{Baines09}
{Baines}, E.~K., {McAlister}, H.~A., {ten Brummelaar}, T.~A., {et~al.} 2009,
  \apj, 701, 154

\bibitem[{{Baines} {et~al.}(2008){Baines}, {McAlister}, {ten Brummelaar},
  {Turner}, {Sturmann}, {Sturmann}, {Goldfinger}, \& {Ridgway}}]{Baines08}
---. 2008, \apj, 680, 728

\bibitem[{{Borucki} {et~al.}(2010){Borucki}, {Koch}, {Basri}, {Batalha},
  {Brown}, {Caldwell}, {Caldwell}, {Christensen-Dalsgaard}, {Cochran},
  {DeVore}, {Dunham}, {Dupree}, {Gautier}, {Geary}, {Gilliland}, {Gould},
  {Howell}, {Jenkins}, {Kondo}, {Latham}, {Marcy}, {Meibom}, {Kjeldsen},
  {Lissauer}, {Monet}, {Morrison}, {Sasselov}, {Tarter}, {Boss}, {Brownlee},
  {Owen}, {Buzasi}, {Charbonneau}, {Doyle}, {Fortney}, {Ford}, {Holman},
  {Seager}, {Steffen}, {Welsh}, {Rowe}, {Anderson}, {Buchhave}, {Ciardi},
  {Walkowicz}, {Sherry}, {Horch}, {Isaacson}, {Everett}, {Fischer}, {Torres},
  {Johnson}, {Endl}, {MacQueen}, {Bryson}, {Dotson}, {Haas}, {Kolodziejczak},
  {Van Cleve}, {Chandrasekaran}, {Twicken}, {Quintana}, {Clarke}, {Allen},
  {Li}, {Wu}, {Tenenbaum}, {Verner}, {Bruhweiler}, {Barnes}, \&
  {Prsa}}]{Borucki10}
{Borucki}, W.~J., {Koch}, D., {Basri}, G., {et~al.} 2010, Science, 327, 977

\bibitem[{{Boyajian} {et~al.}(2013){Boyajian}, {von Braun}, {van Belle},
  {Farrington}, {Schaefer}, {Jones}, {White}, {McAlister}, {ten Brummelaar},
  {Ridgway}, {Gies}, {Sturmann}, {Sturmann}, {Turner}, {Goldfinger}, \&
  {Vargas}}]{Boyajian13}
{Boyajian}, T.~S., {von Braun}, K., {van Belle}, G., {et~al.} 2013, \apj, 771,
  40

\bibitem[{{Brewer} {et~al.}(2015){Brewer}, {Fischer}, {Basu}, {Valenti}, \&
  {Piskunov}}]{Brewer15}
{Brewer}, J.~M., {Fischer}, D.~A., {Basu}, S., {Valenti}, J.~A., \& {Piskunov},
  N. 2015, \apj, 805, 126

\bibitem[{{Brewer} {et~al.}(2016){Brewer}, {Fischer}, {Valenti}, \&
  {Piskunov}}]{Brewer16}
{Brewer}, J.~M., {Fischer}, D.~A., {Valenti}, J.~A., \& {Piskunov}, N. 2016,
  ArXiv e-prints, arXiv:1606.07929

\bibitem[{{Bruntt} {et~al.}(2012){Bruntt}, {Basu}, {Smalley}, {Chaplin},
  {Verner}, {Bedding}, {Catala}, {Gazzano}, {Molenda-{\.Z}akowicz}, {Thygesen},
  {Uytterhoeven}, {Hekker}, {Huber}, {Karoff}, {Mathur}, {Mosser},
  {Appourchaux}, {Campante}, {Elsworth}, {Garc{\'{\i}}a}, {Handberg},
  {Metcalfe}, {Quirion}, {R{\'e}gulo}, {Roxburgh}, {Stello},
  {Christensen-Dalsgaard}, {Kawaler}, {Kjeldsen}, {Morris}, {Quintana}, \&
  {Sanderfer}}]{Bruntt12}
{Bruntt}, H., {Basu}, S., {Smalley}, B., {et~al.} 2012, \mnras, 423, 122

\bibitem[{{Delfosse} {et~al.}(2000){Delfosse}, {Forveille}, {S{\'e}gransan},
  {Beuzit}, {Udry}, {Perrier}, \& {Mayor}}]{Delfosse2000}
{Delfosse}, X., {Forveille}, T., {S{\'e}gransan}, D., {et~al.} 2000, \aap, 364,
  217

\bibitem[{{Dotter} {et~al.}(2008){Dotter}, {Chaboyer}, {Jevremovi{\'c}},
  {Kostov}, {Baron}, \& {Ferguson}}]{Dotter2008}
{Dotter}, A., {Chaboyer}, B., {Jevremovi{\'c}}, D., {et~al.} 2008, \apjs, 178,
  89

\bibitem[{{Gaia Collaboration} {et~al.}(2016){Gaia Collaboration}, {Brown},
  {Vallenari}, {Prusti}, {de Bruijne}, {Mignard}, {Drimmel}, \&
  {co-authors}}]{Gaia16a}
{Gaia Collaboration}, {Brown}, A.~G.~A., {Vallenari}, A., {et~al.} 2016, ArXiv
  e-prints, arXiv:1609.04172

\bibitem[{{Gaidos} {et~al.}(2013){Gaidos}, {Fischer}, {Mann}, \&
  {Howard}}]{Gaidos13}
{Gaidos}, E., {Fischer}, D.~A., {Mann}, A.~W., \& {Howard}, A.~W. 2013, \apj,
  771, 18

\bibitem[{{Gray}(1992)}]{Gray92}
{Gray}, D.~F. 1992, {The observation and analysis of stellar photospheres.}

\bibitem[{{Gray} \& {Corbally}(2009)}]{Gray09}
{Gray}, R.~O., \& {Corbally}, J., C. 2009, {Stellar Spectral Classification}

\bibitem[{{Henry} {et~al.}(2013){Henry}, {Kane}, {Wang}, {Wright}, {Boyajian},
  {von Braun}, {Ciardi}, {Dragomir}, {Farrington}, {Fischer}, {Hinkel},
  {Howard}, {Jensen}, {Laughlin}, {Mahadevan}, \& {Pilyavsky}}]{Henry13}
{Henry}, G.~W., {Kane}, S.~R., {Wang}, S.~X., {et~al.} 2013, \apj, 768, 155

\bibitem[{{Howard} {et~al.}(2010){Howard}, {Johnson}, {Marcy}, {Fischer},
  {Wright}, {Bernat}, {Henry}, {Peek}, {Isaacson}, {Apps}, {Endl}, {Cochran},
  {Valenti}, {Anderson}, \& {Piskunov}}]{Howard10}
{Howard}, A.~W., {Johnson}, J.~A., {Marcy}, G.~W., {et~al.} 2010, \apj, 721,
  1467

\bibitem[{{Huber} {et~al.}(2013){Huber}, {Chaplin}, {Christensen-Dalsgaard},
  {Gilliland}, {Kjeldsen}, {Buchhave}, {Fischer}, {Lissauer}, {Rowe},
  {Sanchis-Ojeda}, {Basu}, {Handberg}, {Hekker}, {Howard}, {Isaacson},
  {Karoff}, {Latham}, {Lund}, {Lundkvist}, {Marcy}, {Miglio}, {Silva Aguirre},
  {Stello}, {Arentoft}, {Barclay}, {Bedding}, {Burke}, {Christiansen},
  {Elsworth}, {Haas}, {Kawaler}, {Metcalfe}, {Mullally}, \&
  {Thompson}}]{Huber13}
{Huber}, D., {Chaplin}, W.~J., {Christensen-Dalsgaard}, J., {et~al.} 2013,
  \apj, 767, 127

\bibitem[{{Kervella} {et~al.}(2003){Kervella}, {Th{\'e}venin}, {Morel},
  {Bord{\'e}}, \& {Di Folco}}]{Kervella03}
{Kervella}, P., {Th{\'e}venin}, F., {Morel}, P., {Bord{\'e}}, P., \& {Di
  Folco}, E. 2003, \aap, 408, 681

\bibitem[{{Kurucz} {et~al.}(1984){Kurucz}, {Furenlid}, {Brault}, \&
  {Testerman}}]{Kurucz84}
{Kurucz}, R.~L., {Furenlid}, I., {Brault}, J., \& {Testerman}, L. 1984, {Solar
  flux atlas from 296 to 1300 nm}

\bibitem[{{Lindegren} {et~al.}(2016){Lindegren}, {Lammers}, {Bastian},
  {Hern{\'a}ndez}, {Klioner}, {Hobbs}, {Bombrun}, {Michalik}, {Ramos-Lerate},
  {Butkevich}, {Comoretto}, {Joliet}, {Holl}, {Hutton}, {Parsons},
  {Steidelm{\"u}ller}, {Abbas}, {Altmann}, {Andrei}, {Anton}, {Bach},
  {Barache}, {Becciani}, {Berthier}, {Bianchi}, {Biermann}, {Bouquillon},
  {Bourda}, {Br{\"u}semeister}, {Bucciarelli}, {Busonero}, {Carlucci},
  {Casta{\~n}eda}, {Charlot}, {Clotet}, {Crosta}, {Davidson}, {de Felice},
  {Drimmel}, {Fabricius}, {Fienga}, {Figueras}, {Fraile}, {Gai}, {Garralda},
  {Geyer}, {Gonz{\'a}lez-Vidal}, {Guerra}, {Hambly}, {Hauser}, {Jordan},
  {Lattanzi}, {Lenhardt}, {Liao}, {L{\"o}ffler}, {McMillan}, {Mignard}, {Mora},
  {Morbidelli}, {Portell}, {Riva}, {Sarasso}, {Serraller}, {Siddiqui}, {Smart},
  {Spagna}, {Stampa}, {Steele}, {Taris}, {Torra}, {van Reeven}, {Vecchiato},
  {Zschocke}, {de Bruijne}, {Gracia}, {Raison}, {Lister}, {Marchant},
  {Messineo}, {Soffel}, {Osorio}, {de Torres}, \& {O'Mullane}}]{Lindegren2016}
{Lindegren}, L., {Lammers}, U., {Bastian}, U., {et~al.} 2016, ArXiv e-prints,
  arXiv:1609.04303

\bibitem[{{Mann} {et~al.}(2013){Mann}, {Brewer}, {Gaidos}, {L{\'e}pine}, \&
  {Hilton}}]{Mann2013}
{Mann}, A.~W., {Brewer}, J.~M., {Gaidos}, E., {L{\'e}pine}, S., \& {Hilton},
  E.~J. 2013, \aj, 145, 52

\bibitem[{{Mann} {et~al.}(2014){Mann}, {Deacon}, {Gaidos}, {Ansdell}, {Brewer},
  {Liu}, {Magnier}, \& {Aller}}]{Mann2014}
{Mann}, A.~W., {Deacon}, N.~R., {Gaidos}, E., {et~al.} 2014, \aj, 147, 160

\bibitem[{{Mann} {et~al.}(2015){Mann}, {Feiden}, {Gaidos}, {Boyajian}, \& {von
  Braun}}]{Mann15}
{Mann}, A.~W., {Feiden}, G.~A., {Gaidos}, E., {Boyajian}, T., \& {von Braun},
  K. 2015, \apj, 804, 64

\bibitem[{{Mann} \& {von Braun}(2015)}]{Mann2015b}
{Mann}, A.~W., \& {von Braun}, K. 2015, \pasp, 127, 102

\bibitem[{{Michalik} {et~al.}(2015){Michalik}, {Lindegren}, \&
  {Hobbs}}]{Michalik2015}
{Michalik}, D., {Lindegren}, L., \& {Hobbs}, D. 2015, \aap, 574, A115

\bibitem[{{Morgan} {et~al.}(1943){Morgan}, {Keenan}, \& {Kellman}}]{Morgan43}
{Morgan}, W.~W., {Keenan}, P.~C., \& {Kellman}, E. 1943, {An atlas of stellar
  spectra, with an outline of spectral classification}

\bibitem[{{Morton}(2015)}]{isochrones}
{Morton}, T.~D. 2015, {isochrones: Stellar model grid package}, Astrophysics
  Source Code Library, , , ascl:1503.010

\bibitem[{Newville {et~al.}(2014)Newville, Stensitzki, Allen, \&
  Ingargiola}]{lmfit}
Newville, M., Stensitzki, T., Allen, D.~B., \& Ingargiola, A. 2014, {LMFIT:
  Non-Linear Least-Square Minimization and Curve-Fitting for Python¶}, , ,
  doi:10.5281/zenodo.11813

\bibitem[{{Perryman} {et~al.}(1997){Perryman}, {Lindegren}, {Kovalevsky},
  {Hoeg}, {Bastian}, {Bernacca}, {Cr{\'e}z{\'e}}, {Donati}, {Grenon},
  {Grewing}, {van Leeuwen}, {van der Marel}, {Mignard}, {Murray}, {Le Poole},
  {Schrijver}, {Turon}, {Arenou}, {Froeschl{\'e}}, \& {Petersen}}]{Perryman97}
{Perryman}, M.~A.~C., {Lindegren}, L., {Kovalevsky}, J., {et~al.} 1997, \aap,
  323, L49

\bibitem[{{Pickles}(1998)}]{Pickles1998}
{Pickles}, A.~J. 1998, \pasp, 110, 863

\bibitem[{{Ricker} {et~al.}(2014){Ricker}, {Winn}, {Vanderspek}, {Latham},
  {Bakos}, {Bean}, {Berta-Thompson}, {Brown}, {Buchhave}, {Butler}, {Butler},
  {Chaplin}, {Charbonneau}, {Christensen-Dalsgaard}, {Clampin}, {Deming},
  {Doty}, {De Lee}, {Dressing}, {Dunham}, {Endl}, {Fressin}, {Ge}, {Henning},
  {Holman}, {Howard}, {Ida}, {Jenkins}, {Jernigan}, {Johnson}, {Kaltenegger},
  {Kawai}, {Kjeldsen}, {Laughlin}, {Levine}, {Lin}, {Lissauer}, {MacQueen},
  {Marcy}, {McCullough}, {Morton}, {Narita}, {Paegert}, {Palle}, {Pepe},
  {Pepper}, {Quirrenbach}, {Rinehart}, {Sasselov}, {Sato}, {Seager},
  {Sozzetti}, {Stassun}, {Sullivan}, {Szentgyorgyi}, {Torres}, {Udry}, \&
  {Villasenor}}]{Ricker14}
{Ricker}, G.~R., {Winn}, J.~N., {Vanderspek}, R., {et~al.} 2014, in \procspie,
  Vol. 9143, Space Telescopes and Instrumentation 2014: Optical, Infrared, and
  Millimeter Wave, 914320

\bibitem[{{Sneden}(1973)}]{Sneden73}
{Sneden}, C.~A. 1973, PhD thesis, The University of Texas at Austin

\bibitem[{{Torres} {et~al.}(2010){Torres}, {Andersen}, \&
  {Gim{\'e}nez}}]{Torres10}
{Torres}, G., {Andersen}, J., \& {Gim{\'e}nez}, A. 2010, \aapr, 18, 67

\bibitem[{{Valenti} \& {Fischer}(2005)}]{Valenti05}
{Valenti}, J.~A., \& {Fischer}, D.~A. 2005, \apjs, 159, 141

\bibitem[{{Valenti} \& {Piskunov}(1996)}]{Valenti96}
{Valenti}, J.~A., \& {Piskunov}, N. 1996, \aaps, 118, 595

\bibitem[{{van Belle} \& {von Braun}(2009)}]{VanBelle09}
{van Belle}, G.~T., \& {von Braun}, K. 2009, \apj, 694, 1085

\bibitem[{{van Leeuwen}(2007)}]{VanLeeuwen07}
{van Leeuwen}, F. 2007, \aap, 474, 653

\bibitem[{{Vogt} {et~al.}(1994){Vogt}, {Allen}, {Bigelow}, {Bresee}, {Brown},
  {Cantrall}, {Conrad}, {Couture}, {Delaney}, {Epps}, {Hilyard}, {Hilyard},
  {Horn}, {Jern}, {Kanto}, {Keane}, {Kibrick}, {Lewis}, {Osborne},
  {Pardeilhan}, {Pfister}, {Ricketts}, {Robinson}, {Stover}, {Tucker}, {Ward},
  \& {Wei}}]{Vogt94}
{Vogt}, S.~S., {Allen}, S.~L., {Bigelow}, B.~C., {et~al.} 1994, in \procspie,
  Vol. 2198, Instrumentation in Astronomy VIII, ed. D.~L. {Crawford} \& E.~R.
  {Craine}, 362

\bibitem[{{von Braun} {et~al.}(2011{\natexlab{a}}){von Braun}, {Boyajian}, {ten
  Brummelaar}, {Kane}, {van Belle}, {Ciardi}, {Raymond}, {L{\'o}pez-Morales},
  {McAlister}, {Schaefer}, {Ridgway}, {Sturmann}, {Sturmann}, {White},
  {Turner}, {Farrington}, \& {Goldfinger}}]{VonBraun11a}
{von Braun}, K., {Boyajian}, T.~S., {ten Brummelaar}, T.~A., {et~al.}
  2011{\natexlab{a}}, \apj, 740, 49

\bibitem[{{von Braun} {et~al.}(2011{\natexlab{b}}){von Braun}, {Boyajian},
  {Kane}, {van Belle}, {Ciardi}, {L{\'o}pez-Morales}, {McAlister}, {Henry},
  {Jao}, {Riedel}, {Subasavage}, {Schaefer}, {ten Brummelaar}, {Ridgway},
  {Sturmann}, {Sturmann}, {Mazingue}, {Turner}, {Farrington}, {Goldfinger}, \&
  {Boden}}]{VonBraun11b}
{von Braun}, K., {Boyajian}, T.~S., {Kane}, S.~R., {et~al.} 2011{\natexlab{b}},
  \apjl, 729, L26

\bibitem[{{von Braun} {et~al.}(2012){von Braun}, {Boyajian}, {Kane}, {Hebb},
  {van Belle}, {Farrington}, {Ciardi}, {Knutson}, {ten Brummelaar},
  {L{\'o}pez-Morales}, {McAlister}, {Schaefer}, {Ridgway}, {Collier Cameron},
  {Goldfinger}, {Turner}, {Sturmann}, \& {Sturmann}}]{VonBraun12}
---. 2012, \apj, 753, 171

\bibitem[{{von Braun} {et~al.}(2014){von Braun}, {Boyajian}, {van Belle},
  {Kane}, {Jones}, {Farrington}, {Schaefer}, {Vargas}, {Scott}, {ten
  Brummelaar}, {Kephart}, {Gies}, {Ciardi}, {L{\'o}pez-Morales}, {Mazingue},
  {McAlister}, {Ridgway}, {Goldfinger}, {Turner}, \& {Sturmann}}]{VonBraun14}
{von Braun}, K., {Boyajian}, T.~S., {van Belle}, G.~T., {et~al.} 2014, \mnras,
  438, 2413

\end{thebibliography}

\end{document}